\documentclass [12pt]{article}
\usepackage{lscape}                                           %
\usepackage[centertags]{amsmath}
\usepackage{amsfonts} \usepackage{amssymb}\usepackage{eufrak}
\usepackage{colordvi}
\usepackage{epsfig}
\usepackage{a4}
\usepackage{cite}
\usepackage{lscape}
\newcommand{\dslash}{\ensuremath \slash\hspace{-0.25cm}}
\newcommand{\COMMENTO}[1]{}

\begin{document}
\begin{titlepage}
\rightline{DSF-25-2008}
\rightline{NORDITA-2008-47} \vskip 3.0cm
\centerline{\LARGE \bf  K{\"{a}}hler Metrics and Yukawa Couplings
}\vskip .5cm
\centerline{\LARGE \bf  in Magnetized Brane Models}
\vskip 1.0cm \centerline{\bf P. Di Vecchia$^{a,b}$, A. Liccardo$^{c,d}$, R. Marotta$^d$
and F. Pezzella$^d$}
\vskip .6cm \centerline{\sl $^a$
The Niels Bohr Institute, Blegdamsvej 17, DK-2100 Copenhagen \O, Denmark}
\vskip .4cm
\centerline{\sl $^b$
Nordita, Roslagstullsbacken 23, SE-10691 Stockholm, Sweden}
\vskip .4cm \centerline{\sl
 $^c$ Dipartimento di
Scienze Fisiche, Universit\`a degli Studi ``Federico II'' di Napoli}
\centerline{\sl Complesso Universitario Monte
S. Angelo ed. 6, via Cintia,  I-80126 Napoli, Italy}
\vskip .4cm
\centerline{\sl
$^d$ Istituto Nazionale di Fisica Nucleare, Sezione di Napoli}
\centerline{ \sl Complesso Universitario Monte
S. Angelo ed. 6, via Cintia,  I-80126 Napoli, Italy}

 \vskip 1cm

\begin{abstract}
Using the field theoretical approach introduced by Cremades, Ib{\'{a}}{\~{n}}ez
and Marchesano for describing open strings  attached
to D9 branes having different magnetizations, we  give a
procedure for  determining  the K{\"{a}}hler metrics
of those  open strings in toroidal compactifications.
\end{abstract}
\vfill  {\small{ Work partially supported by the European
Community's Human Potential Programme under contract
MRTN-CT-2004-005104 ``Constituents, Fundamental Forces and Symmetries
of the Universe''. }}
\end{titlepage}

\newpage

\tableofcontents       %
\vskip 1cm             %

\section{Introduction}
\label{intro}

In order to  connect  string theory with experimental data  we need to
develop techniques that allow us to compute the low-energy
four-dimensional effective action  that can be directly compared
with what we  observe in high energy
experiments, starting from the original ten dimensional theory
compactified on $R^{3,1} \times M_6$, where $M_6$ is a compact six
dimensional manifold. In particular, it is important to understand
how the four-dimensional physics depends on the size and the shape
of  $M_6$.

Recently, in the framework of toroidal compactifications with a
number of stacks of intersecting or of their T-dual magnetized D
branes, semi-realistic string extensions of  the Standard Model and
of the Minimal Supersymmetric Standard Model have been
constructed~\footnote{See Ref.~\cite{0307262} for a description
of some of those models.}. In the magnetized brane scenario
quarks and leptons correspond to open strings  having their
end-points attached to D branes with different magnetizations.
Those open strings are  called in the literature {\em dycharged, chiral}
or {\em twisted} strings. Since in the case of a constant magnetization
the dynamics of the open strings can be completely and
analytically determined and the open strings can be exactly
quantized~\cite{BP9209}, from the computation of string amplitudes
one can in principle determine the low-energy four-dimensional
effective Lagrangian involving those fermionic chiral strings and
their supersymmetric bosonic partners.  In particular, by
computing three and four-point string amplitudes, in
Refs.~\cite{0404134,0512067}  the dependence on the magnetization
of the K{\"{a}}hler metrics of the twisted strings has been
determined. On the other hand, using instanton calculus and the
holomorphicity of the superpotential for $M_6 = T^2 \times T^2
\times T^2$, it has been seen~\cite{0705.2366,0709.0245,0711.0866}
that the  K{\"{a}}hler metrics of the twisted strings contain an
additional explicit dependence on the moduli and also an arbitrary
factor that up to now has not been possible to fix by an
explicit string calculation.

However, if one is not immediately interested in the string corrections to the
parameters of the low-energy four-dimensional effective action, one does not
really need to compute string amplitudes, but
one can directly start from the action of ${\cal{N}}=1$ super Yang-Mills
in ten dimensions with gauge group $U(M)$,
that describes  the open strings attached to $M$ D9 branes,  perform
on it a Kaluza-Klein reduction from ten to four dimensions and
derive the low-energy
four-dimensional effective action for the massless excitations.
In particular, by imposing that the background gauge field in the six compact
dimensions   and along  the Cartan subalgebra of $U(M)$ is non-vanishing and
corresponds to a constant gauge field strength, one gets
a field theoretical description of the twisted open strings, previously defined.
This approach pioneered in a beautiful paper  by Cremades, Ib{\'{a}}{\~{n}}ez
and Marchesano~\cite{0404229} for computing the Yukawa couplings
of chiral matter and extended in Ref.~\cite{0806.4748} to orbifolds of toroidal models
and in Ref.~\cite{0807.0789} to some non-toroidal compactifications,  is
the one that we are going to use for computing the K{\"{a}}hler
metrics of twisted open strings.

In this paper, following  the approach of Ref.~\cite{0404229},
we  give a procedure for computing the K{\"{a}}hler metric
of  twisted and untwisted scalar fields. In fact, unlike Ref.~\cite{0404229},
we  do not normalize to 1 the wave-functions in the compact
extra-dimensions, but instead
we keep  the moduli dependence that naturally comes from the
integral over the compact manifold.
In this way we can correctly
reproduce the dependence on the moduli of the K{\"{a}}hler metrics
apart from factors involving the magnetizations. In order to get
also these factors,  we add a normalization factor for each scalar field and for
its supersymmetric fermionic partner that is then
determined by requiring that
the Yukawa couplings, that we also compute, come from a
holomorphic  super-potential.  In this way we reproduce  the K{\"{a}}hler
metric of the adjoint scalars, of those of the hypermultiplet and of those of
the  chiral multiplet without additional arbitrary factors.
It must also be said, however, that this is of course the minimal way to eliminate
non-holomorphic factors in the Yukawa couplings. One could, in principle,
also include factors that do not spoil the holomorphicity. Finally, there is also
the issue of how the four-dimensional  complex field depends on the
ten-dimensional fields that will be discussed in the conclusions.

The paper is organized as follows. In Sect. \ref{2} we consider the terms
of the ten-dimensional action that are relevant in our calculations and we
perform the Kaluza-Klein reduction from ten to four dimensions.
Using the wave functions in the extra dimensions, that are computed in
Appendix \ref{B},  in Sect. \ref{Kaehler}
we determine the K\"{a}hler metrics of the various scalar fields apart from
a normalization factor for each four-dimensional field.  In Sect. \ref{YUKA}
we compute the Yukawa couplings and fix the previous
normalization factors by requiring that the Yukawa couplings come from a
holomorphic superpotential. In Sect. \ref{FIX} we insert those normalization
factors  in the two-point functions for the various scalar fields and we determine
their K{\"{a}}hler metrics showing that the expression obtained are consistent
with previous string calculations in the field theory limit.  Sect. \ref{conclu}
is devoted to the conclusions and to a discussion of the form of
the four-dimensional
scalar fields in terms of the original ten-dimensional ones.

Some Appendices follow. Appendix  \ref{T2} is devoted to a description
of the torus $T^2$ and of the moduli used in supergravity. In Appendix \ref{B}
we solve both the bosonic and fermionic
eigenvalue equations for the wave functions in the extra
dimensions obtaining the explicit wave functions. In Appendix  \ref{C} we add
few details on the calculation of the Yukawa couplings and finally in
Appendix \ref{susy}
we discuss the four-dimensional supersymmetry transformations.

\section{The KK reduction of the relevant terms of  the action}
\label{2}

The starting point of our analysis is the low-energy limit of the DBI
action describing a
set of  $M$ D9 branes, namely supersymmetric ${\cal{N}}=1$
super Yang-Mills with gauge group $U(M)$:
\begin{eqnarray}
S = \frac{1}{g^2} \int d^{10} X {\rm Tr}
\bigg(- { 1 \over 4} F_{MN} F^{MN}  +
{i \over 2}\bar{\lambda} \Gamma^{M} {D}_{M} \lambda \bigg),
\label{10dimla}
\end{eqnarray}
where $g^2 =  4 \pi {\rm e}^{\phi_{10}} (2 \pi \sqrt{\alpha'})^6$ and
\begin{eqnarray}
F_{MN} = \nabla_M A_N - \nabla_N A_M  - i[A_M, A_N]~~;~~
{D}_M \lambda = \nabla_M \psi - i [A_M,\lambda]
\label{FMN}
\end{eqnarray}
being $ \lambda$ a ten dimensional Weyl-Majorana spinor.

We separate the generators of the gauge group
into those, called $U_a$, that are in
the Cartan subalgebra and those, called  $e_{ab}$,  outside of
it \cite{0404229} \cite{0807.0789}:
\begin{eqnarray}
(U_a)_{ij} = \delta_{ai} \delta_{aj}, \qquad (e_{ab})_{ij} =
\delta_{ai} \delta_{bj}  \quad (a\neq b).
\label{not62}
\end{eqnarray}
The gauge field $A_M$ and the gaugino are expanded as
\begin{eqnarray}
A_M = B_M + W_M = B_M^a U_a + W_M^{ab} e_{ab}~~;~~
\lambda  =  \chi + \Psi = \chi^{a}U_a + \Psi^{ab}e_{ab}.
\label{exp98}
\end{eqnarray}
Requiring that $A_M^{\dagger} = A_M$ implies
that $B_M^a$ is real and $(W_M^{ab})^{*} = W_{{M}}^{ba}$.
The same is true for the gaugino and its components $ \chi$ and $\Psi$.

By inserting in Eq. (\ref{10dimla}) the expansions given  in Eq. (\ref{exp98}),
we can rewrite the original action in terms of the fields $B$, $W$, $\chi$ and $\Psi$.
Its explicit expression can be found in Refs.~\cite{0404229} \cite{0807.0789}.

We  separate the ten-dimensional coordinate
$X^M$ into a four-dimensional non-compact coordinate $x^\mu$
and a six-dimensional compact variable $y^i$ and perform a
Kaluza-Klein reduction of the Lagrangian in Eq. (\ref{10dimla})
expanding around the background fields:
\begin{eqnarray}
B_M^{a}(x^{\mu} , y^{i} ) & = & \langle B_M^{a} \rangle (y^i)+
 \delta B_M^{a}(x^{\mu} , y^{i}), \\
W_M^{ab}(x^{\mu} , y^{i}) & = & 0 + \Phi_M^{ab}(x^{\mu} , y^{i})
\label{expa3}
\end{eqnarray}
where, in order to keep the four-dimensional Lorentz invariance, we
allow a non-vanishing background value  $\langle B_M^{a}
\rangle(y^i)$ only for $M= i$, i.e. along the compact extra-dimensions. The presence of different background values along the Cartan
subalgebra breaks the original $U(M)$ symmetry into $(U(1))^M$. In
terms of D branes this corresponds to generate $M$ stacks, each
consisting of one D brane with its own magnetization, different from that of the other branes, and the
fields  $ \Phi_M^{ab}(x^{\mu} , y^{i})$ for $M=i$ describe twisted
open strings with the two end-points attached respectively
to two  D branes $a$ and  $b$
having different magnetizations. If some of the background values are equal,
then the original gauge group $U(M)$  is broken into a product of non-abelian
subgroups.

In the following we will not rewrite the entire action in terms of  the fields
introduced above, but we will only write the relevant terms, namely the quadratic
terms involving the scalar and fermion fields and the trilinear terms involving a scalar and two fermions: we will derive the K{\"{a}}hler metrics from the former
and  the Yukawa couplings from the latter. We will also restrict our considerations to  toroidal compactifications.

The quadratic terms for the
fields $ \Phi_M^{ab}(x^{\mu} , y^{i})$ are the following:
\begin{eqnarray}
S_{2}^{(\Phi)} =
\frac{1}{2g^2} \int d^4 x \sqrt{G_4} \int d^6 y \sqrt{G_6}   \Phi^{j ba}
\left[   G^{i}_{\,\,j}  \left( D_{\mu}  D^{\mu}      +  {\tilde{D}}_{k}
{\tilde{D}}^{k} \right)  + 2 i   \langle (F_{B})^{i}_{\,\,j} \rangle^{ab}
\right]  \Phi_{i}^{ab}
\label{quacsca1}
\end{eqnarray}
where
\begin{eqnarray}
 {D}_\mu \Phi^{ab}_{j}= \partial_{\mu} \Phi^{ab}_{j}-i (B_{\mu}^{a} -B_{\mu}^{b} ) \Phi^{ab}_{j} ~~;~~
 \tilde{D}_i \Phi^{ab}_{j}= \partial_i \Phi^{ab}_{j} - i  \left(  \langle B_i^a \rangle  -  \langle  B_i^b \rangle \right)  \Phi^{ab}_{j}
\label{covder11}
\end{eqnarray}
with
\begin{eqnarray}
<(F_{B})^{i}_{j} >^{ab} \,\, \equiv(F_{B}^{a})^{i}_{\,\,\,j} - (F_{B}^{b})^{i}_{\,\,\,j}
\label{F-F}
\end{eqnarray}
where $(F_{B}^{a})^{i}_{\,\,j}$  is the field strength obtained from
the background field $B^a$.
Analogously we can consider the quadratic term for the
fields $\delta B^{a}_{i} ( x^{\mu} , y^i )$ obtaining:
\begin{eqnarray}
S_{2}^{(\delta B)} = \frac{1}{2 g^2} \int d^4 x \sqrt{G_4} \int d^6 y \sqrt{G_6}
\delta B _{i }^{a} \left(\partial_j \partial^{j}  + D_{\mu} D^{\mu}  \right) \delta B^{ai}
\label{adjsca}
\end{eqnarray}
where the gauge $ \partial_{M} \delta B^{aM} =0 $ has been chosen.

The quadratic term of the fermions $\Psi^{ab} (x^{\mu}, y^i )$  is given by:
\begin{eqnarray}
{ S}_{2}^{(\Psi)}  =  {i \over 2g^2}   \int d^4 x \sqrt{G_4} \int d^6 y \sqrt{G_6}
\bar{\Psi}^{ba} \left( \Gamma^\mu
{D}_\mu      +    \Gamma^i \tilde{D}_i \right) \Psi^{ab}
\label{LF2b}
\end{eqnarray}
where ${D}_\mu\Psi^{ab}$ and $ \tilde{D}_i \Psi^{ab}$ are the same as in Eq.
(\ref{covder11}).

The trilinear Yukawa couplings are given by:
\begin{eqnarray}
S_{3}^{(\Phi)} = {1 \over 2 g^2} \int d^4 x \sqrt{G_4} \int d^6 y \sqrt{G_6}
\left(\bar{\Psi}^{ca}\Gamma^i
\Phi_{i}^{ab}\Psi^{bc}-\bar{\Psi}^{ca}\Gamma^i\Phi_{i}^{bc}
\Psi^{ab}\right)
\label{LYb}
\end{eqnarray}
and by
\begin{eqnarray}
{ S}_{3}^{(\delta B)} = \frac{1}{2g^2}
\int d^4 x \sqrt{G_4} \int d^6 y \sqrt{G_6}
 \bar{\Psi}^{ab}( {\dslash{\delta B}}^b-{\dslash{\delta B}}^a)\Psi^{ba}
\label{LYbc}
\end{eqnarray}
respectively for the twisted scalar $\Phi$ and for the untwisted scalar $\delta B$.

The four-dimensional effective action, corresponding to the
ten-dimensional actions given above, is obtained by
expanding the ten-dimensional fields as follows~\footnote{The wave-function
in the extra dimensions can in principle also depend on the index $i$, but this does not happen in our case as one can see from Eqs.  (\ref{eq88}).}:
\begin{eqnarray}
\Phi^{ab}_{i} (X) = \sum_n \varphi^{ab}_{n,i} (x^{\mu}  )  \phi^{ab}_{n} (y^{i} )
~~;~~\Psi^{ab} (X) =  \sum_n \psi_n^{ab}(x^{\mu}) \otimes  \eta_n^{ab}(y^i)\,\,.
\label{exps63}
\end{eqnarray}
The spectrum of the Kaluza-Klein states and their
wave functions along the compact directions are
obtained by  solving the eigenvalue equations for the six-dimensional
Laplace and Dirac operators:
\begin{eqnarray}
-{\tilde{D}}_{k}  {\tilde{D}}^{k}  \phi^{ab}_{n}  =
 m_{n}^{2} \phi^{ab}_{n}
 ~~,~~
i \gamma_{(6)}^i \tilde{D}_i \eta^{ab}_n =
\lambda_n\,\eta^{ab}_n
\label{eq88}
\end{eqnarray}
with the correct periodicity conditions along the compactified directions.
It is worthwhile to remind that, according to a  standard procedure followed in dimensionally reducing the ten-dimensional Dirac equation \cite{GSW}, it is necessary to
make the operators $\Gamma^{\mu}D_{\mu}$  and $\Gamma^{i} \tilde{D}_{i}$ commute in order to properly define simultaneous eigenstates. This is accomplished by multiplying the latter operators with $\Gamma^{(5)} = i \Gamma^{0} \Gamma^{1} \Gamma^{2} \Gamma^{3}$, which yields to Eq. (\ref{eq88}) after
having used the decomposition:
\begin{eqnarray}
{\Gamma}^\mu= \gamma^\mu_{(4)}\otimes \mathbb{I}_{(6)}~~,~~
{\Gamma}^i=\gamma^5_{(4)}\otimes\gamma^i_{(6)} ~~.~~
\label{decoDirac}
\end{eqnarray}
Inserting Eq. (\ref{exps63}) and the first equation in (\ref{eq88})
in Eq. (\ref{quacsca1}) and
using the coordinates $z$ and ${\bar{z}}$  introduced in Eq. (\ref{comcor})
of Appendix A
for describing  the torus $T^2$, one gets~\footnote{In this
formula the indices $i$ and $j$ run over each of the three torus $T^2$.}:
\begin{eqnarray}
S_2^{(\Phi)}&=&\frac{1}{2g^2} \int d^4 x \sqrt{G_4}
\prod_{r=1}^{3} \left[ (2 \pi R)^2 \int d^2 z_r
\sqrt{G^r}\right]   \nonumber \\
&& \times   \sum_{r=1}^{3}
\Phi^{j \,\,ba}_{ r}   \left[  G_{j}^{r\,\,i} \left( D_{\mu}  D^{\mu}      - m_{n}^{2}    \right)
+ 2 i   \frac{<(F_{r})^{\,\,i}_{j}>^{ab}}{(2 \pi R)^2}      \right]  \Phi^{ab  }_{ri}
\label{la62}
\end{eqnarray}
that, more explicitly, can be written as:
\begin{eqnarray}
S_2^{(\Phi)}&=& \frac{1}{2g^2} \int d^4 x \sqrt{G_4}   \sum_{n} \prod_{r=1}^{3} \left[
(2 \pi R)^2 \int d^2 z_r
\sqrt{G^r}    \right]  \phi_{n}^{ba }   \phi^{ab}_{n}
\nonumber \\
&&\times
\left\{\sum_{r=1}^{3} N_{\varphi_r}^{2} \left[ \varphi_{nr}^{ba,z} (x)
\left[  D_{\mu}  D^{\mu} - m_{n}^{2}    +
 \frac{4 \pi I_r}{( 2 \pi R)^2 {\cal{T}}_{2}^{(r)}}    \right]
 \varphi_{nr z}^{ab} (x) \right]
  \right.  \nonumber \\
&&+ \left.
 \sum_{r=1}^{3} N_{\varphi_r}^{2} \left[ \varphi_{nr}^{ba,{\bar{z}}} (x)
\left[  D_{\mu}  D^{\mu} - m_{n}^{2}    -
\frac{4 \pi I_r}{( 2 \pi R)^2 {\cal{T}}_{2}^{(r)}}     \right]
\varphi_{nr {\bar{z}}}^{ab} (x) \right]
 \right\}
\label{lagra391}
\end{eqnarray}
where we have used Eqs. (\ref{Gzz63}) and (\ref{GG564}), in which the first Chern class $I_r$ appears. Moreover,
we have introduced a normalization factor $N_{\varphi_r}$, that in general
will depend on the moduli.  This factor has been fixed in Ref.~\cite{0404229}
requiring that the quadratic terms are canonically normalized.
In this paper we adopt a different procedure and we will fix it
later on  by requiring the holomorphicity of the superpotential.

From Eq. (\ref{lagra391}) we see that there are
 two towers  of Kaluza-Klein states
for each torus, with masses given by:
\begin{eqnarray}
(M_{n, r}^{\pm})^{2} = m_{n}^{2} \pm  \frac{4 \pi I_r}{(2 \pi R)^2
{\cal{T}}_{2}^{(r)}} =
\frac{1}{(2 \pi R)^2} \left[\sum_{s=1}^{3}
\frac{2 \pi |I_{s}| }{{\cal{T}}_{2}^{(s)}}
\left( 2N_s +1 \right) \pm
\frac{4 \pi I_r}{{\cal{T}}_{2}^{(r)}} \right]
\label{KKmass}
\end{eqnarray}
where $N_s$ is an integer given by the oscillator number operator. The presence of
the oscillator number is a consequence of the fact that, as shown in Eq. (\ref{DD96}),
the Laplace operator can be written in terms of the creation and annihilation operators
of an harmonic oscillator.
Notice that, since we use a dimensionless ${\cal{T}}_2$, the factor
$\frac{1}{(2 \pi R)^2}$  in front
is  just there to cancel the dependence of the physical masses
on the unphysical parameter $R$.
One can have a massless state only if the following
condition is satisfied for $I_r>0$ or $I_r<0$~\footnote{This condition is the
field theory limit of the relation that one imposes in string theory in the twisted
sector to keep ${\cal{N}}=1$ supersymmetry.}:
\begin{eqnarray}
\sum_{s=1}^{3} \frac{2 \pi |I_{s}| }{{\cal{T}}_{2}^{(s)}}   -
\frac{4 \pi |I_r| }{{\cal{T}}_{2}^{(r)}}  =0 \Longrightarrow
\frac{1}{2} \sum_{s=1}^{3} \frac{ |I_{s}|}{{\cal{T}}_{2}^{(s)}}   -
\frac{ |I_r|}{{\cal{T}}_{2}^{(r)}}  =0 \,\,.
\label{zerom}
\end{eqnarray}
In this case one  keeps ${\cal{N}}=1$ supersymmetry because there is a massless
scalar that is in the same  chiral multiplet as a fermion that
we will study later. If one of the $I_r$'s is vanishing and the other two are equal, then
we have an additional massless excitation corresponding to  an extended
${\cal{N}}=2$ supersymmetry.

It is convenient to use  fields $\varphi^{I}$  with flat
indices~\footnote{We will discuss in the conclusions the reason of this choice.}:
\begin{eqnarray}
&&\varphi_{{\bar{z}}}^{ab}  = G_{\bar{z}z}
e^{z}_{~I} (\varphi^I )^{ab} \equiv
\sqrt{  \frac{{\cal{T}}_2}{2U_2}}\varphi_{+}^{ab}~~;~~
(\varphi^{\bar{z}})^{ba}  =  e^{\bar{z}}_{~I}(\varphi^I)^{ba}
\equiv  \sqrt{  \frac{2 U_2}{{\cal{T}}_2}}(\varphi_{+}^{ab} )^\dag\nonumber\\
&&\varphi_{{{z}}}^{ab}  = G_{z\bar{z}}
e^{\bar{z}}_{~I} (\varphi^I )^{ab}\equiv
\sqrt{  \frac{{\cal{T}}_2}{2U_2}}\varphi_{-}^{ab}~~;~~
(\varphi^{{z}})^{ba}  =  e^{{z}}_{~I} (\varphi^I )^{ba}
\equiv  \sqrt{  \frac{ 2U_2}{{\cal{T}}_2}} (\varphi_{-}^{ab} )^\dag\nonumber\\
\label{varphi+-}
\end{eqnarray}
where
\begin{eqnarray}
 (\varphi_{+})^{ab}  = \left( \frac{ \varphi^1+i\,\varphi^2}{\sqrt{2}}\right)^{ab}~~;~~
( \varphi_{-})^{ab}  = \left( \frac{ \varphi^1-i\,\varphi^2}{\sqrt{2}}\right)^{ab} ~~;~~
 \varphi_{+}^{ba}=(\varphi_{-}^{ab} )^\dag \,\,.
\label{+-87}
\end{eqnarray}
The  action for the twisted scalars restricted to the lowest modes  of the two
towers of Kaluza-Klein states becomes:
\begin{eqnarray}
S_2^{(\Phi_0)}&=&- \frac{1}{2g^2}  \prod_{s=1}^{3} \left[(2 \pi R)^2 \int d^2 z_s
\sqrt{G^{(z_s , {\bar{z}}_s)} } \right]
  (\phi_{0}^{ab})^{\dagger}(\phi_{0}^{ab})
  \int d^4 x \sqrt{G_4}\sum_{r=1}^3  N_{\varphi_r}^{2}\nonumber\\
&\times & \left[
(D_{\mu} (\varphi^{ab}_{r, +})^{\dag}  (x) )
( D^{\mu} \varphi^{ab}_ {r,+} (x) ) +
(M_{0,r}^+)^2 (\varphi^{ab}_{r,+})^{\dag} (x)
\varphi^{ab}_ {r,+} (x)   \right. \nonumber\\
&+&\left.
(D_{\mu} (\varphi^{ab}_{r,-})^{\dagger} (x) )
( D^{\mu} \varphi^{ab}_ {r,-} (x) ) +(M_{0,r}^-)^2 (\varphi^{ab}_{r,-})^{\dag} (x)
\varphi^{ab}_ {r,-} (x)\right] \,\,.
\label{masca71}
\end{eqnarray}
The susy conditions given in Eq. (\ref{zerom}) show that only one
of the two scalars is massless. In particular,  by choosing in such
equation $r=1$ and
$I_{1}>0$, we see that $\varphi_{1,-}$ is the massless scalar.
The corresponding  internal wave-function has been determined in
Ref.~\cite{0404229} and is the product of three eigenfunctions
\begin{eqnarray}
\phi^{ab}_0=\prod_{r=1}^3\phi_{r,{sign}I_{r}}^{ab;n^r}
\end{eqnarray}
where
\begin{eqnarray}
\phi_{r,+}^{ab;n_r}&=&
e^{\pi iI_r {{z}}_r \frac{ {\rm Im}z_r}{ {\rm Im} U^{(r)} }}\,
\Theta\left[\begin{array}{cc} \frac{2n_r}{ I_r}\\0\end{array}\right]
(I_r  z_r | I_r  U^{(r)}) \qquad {\rm for}~I_r>0 \nonumber\\
\phi_{r,-}^{ab;n_r}&=&
e^{i\pi |I_r| {\bar z}_r \frac{{\rm Im}{\bar z}_r}{{\rm Im} U^{(r)} }}\,
\Theta\left[\begin{array}{cc} \frac{-2n_r}{ I_r}\\0\end{array}\right]
(I_r  {\bar z}_r | I_r {\bar U}^{(r)}) \qquad {\rm for}~I_r<0
\label{wavefu6}
\end{eqnarray}
with
$n_r =0, \dots,  |I_r|-1$ labelling the Landau levels.
Instead,  by taking $r=1$ and $I_{1}<0$ we have that $\varphi_{1,+}$ becomes the massless mode.
It is useful to notice that $(\phi_{r,+}^{ab;n_r})^\dag=\phi_{r,-}^{ba;n_r}$, and
furthermore, the reality of the scalar action implies:
\begin{eqnarray}
\phi^{ba }_{0} = ( \phi_{0}^{ab})^{*}.
\label{comco8}
\end{eqnarray}
In conclusion, by performing the Kaluza-Klein reduction of the low-energy
world-volume action of a stack of D9 branes on
$R^{3,1} \times T^2 \times T^2 \times T^2$, we have found two  towers
of Kaluza-Klein states for each of  the scalar fields $\varphi_{r, \pm}$
for $r=1,2,3$ corresponding
to twisted or dycharged strings. In general,  only the lowest state of one
of the two towers and for a particular value of  $r$ (say $r=1$ if Eq. (\ref{zerom})
is satisfied for $r=1$) is
massless,  depending on the sign of $I_{1}$.
We have now all the elements for computing the
K{\"{a}}hler metric of the scalars $\varphi_{\pm}$. This will be done in
Sect. \ref{Kaehler}.

Next  we consider Eq. (\ref{adjsca}) for the adjoint scalars
and expand  the fluctuations as follows:
\begin{eqnarray}
\delta B^{a}_{i} (x^{\mu} , y^{i} )  = \sum_n C^{a}_{ni} (x^{\mu})
c^{a}_{n } (y^i) \,\,.
\label{kkre4}
\end{eqnarray}
Inserting this expansion in Eq. (\ref{adjsca}) and limiting ourselves to the
constant zero mode we get~\footnote{Also here, as in Eq. (\ref{la62}), with an abuse
of notation, we take the indices $i$ and $j$ running over each of the three tori. Furthermore we let the subindex $0$ drop.}:
\begin{eqnarray}
S^{(\delta B)}_{2} = \frac{1}{2 g^2} \int d^4 x \sqrt{G_4}  \prod_{s=1}^{3} \left[ (2 \pi R)^2
\int d^2 z_s  \sqrt{G^{(z_s , {\bar{z}}_s)} } \right]  \sum_{r=1}^{3} C^{ai }_{ r} (x) \left( \partial_{\mu} \partial^{\mu} + \partial_{j} \partial^{j}    \right) C^{a}_{ri} (x)
\nonumber
\end{eqnarray}
that is equal to:
\begin{eqnarray}
S^{(\delta B)}_{2}= - \frac{1 }{2\cdot 4 \pi  } \, {\rm e}^{- \phi_{10}} \,\prod_{s=1}^{3}
        T_{2}^{(s)} \int d^4 x \sqrt{G_4}
\left[  \sum_{r=1}^{3} G^{ij}_{r} \partial^{\mu} C^{a}_{ri} (x)
\partial_{\mu}
          C^{a}_{rj} (x)  \right]
\label{S2d}
\end{eqnarray}
where we have taken  the constant lowest eigenfunction $c^{a} (y^i) = 1$.

Using the first metric in Eq. (\ref{meGG}) and going to the Einstein frame, we
get the following final expression:
\begin{eqnarray}
S^{(\delta B)}_{2} =
 -\frac{1}{2 } \, {\rm e}^{-\phi_{10}} \,
        T_{2}^{(1)} T_{2}^{(2)}   T_{2}^{(3)} {\rm e}^{2 \phi_4}  \int d^4 x \sqrt{G_4}
\left[  \sum_{r=1}^{3} \frac{1}{T_{2}^{(r)} U_{2}^{(r)}}
\partial^{\mu} {\bar{\varphi}}^{a}_{r}  (x)
\partial_{\mu} \varphi^{a}_{r}  (x)
           \right]
\label{meijbc}
\end{eqnarray}
where
\begin{eqnarray}
\varphi^{a}_{r} \equiv   i\frac{ {\bar{U}}  \tilde{C}^{a}_{r 1} -  \tilde{C}^{a}_{r 2}}{\sqrt{4 \pi}}\,\,.
\label{varphir}
\end{eqnarray}
Here, the  fields $\tilde{C}_r$'s are the ones defined in the $\tilde{x}$ coordinate system introduced in Appendix \ref{T2}.

We finally consider the kinetic term for the twisted fermions. It is obtained by plugging in Eq. (\ref{LF2b}) the Kaluza-Klein mode expansion given in (\ref{exps63}), getting:
\begin{eqnarray}
S_F^{(2)}= \frac{1}{2 g^2}\sum_{n,m}
\int d^4 x \sqrt{G^{(4)}} \left[
\bar{\psi}^{ba}_n \left( \, i \gamma_{(4)}^\mu {D}_\mu \,
+   \lambda_n \gamma_{(4)}^{5}
\right)\psi^{ab}_m\right]
\int d^6 y  \sqrt{G^{(6)}} (\eta^{ab}_n)^\dag \,\eta^{ab}_m
\label{43b}
\end{eqnarray}
where we have used the identity $\bar{\eta}^{ba}=(\eta^{ab})^\dag$ which follows from the structure of  $\Gamma^0$  given in  Eq. (\ref{decoDirac}).

\section{The K{\"{a}}hler metrics}
\label{Kaehler}

In this section we continue the calculation previously started  for
determining the K{\"{a}}hler metric of the scalars $\varphi_{\pm}$. In particular, we pay our attention on the term containing the massless
scalar that we name  $\varphi$. The
K{\"{a}}hler metric $Z$ can be read from the kinetic term for the field
$\varphi$  given by:
\begin{eqnarray}
-  \int d^4 x \sqrt{G_4} \,\,Z( m , {\bar{m}} ) \,
(D_{\mu} {\bar{\varphi}}  (x) )
( D^{\mu} \varphi  (x) )
\label{Kaeme}
\end{eqnarray}
written in the Einstein frame. The field $\varphi$  is related to the fields $\varphi_{-}$
by: $\varphi = \frac{\varphi_{-}}{\sqrt{4 \pi}}$~\footnote{We discuss only the case $I>0$. The final relations are trivially extended to the case $I<0$.}, absorbing in the definition
of the field  the factor  $4 \pi$ present in $g^2$. $m$ and ${\bar{m}}$ stand
for the  moduli.

By comparing this equation with Eq. (\ref{masca71})
the following expression for $Z$ can be obtained:
\begin{eqnarray}
Z ( m , {\bar{m}}) = \frac{{4 \pi \rm e}^{2\phi_4} }{2 g^2} N_{\varphi}^{2}
\prod_{s=1}^{3}
\left[ (2 \pi R)^2 \int d^2 z_s
\sqrt{G^{(z_s , {\bar{z}}_s)} } \right]   \phi_{0}^{ba}
\phi^{ab }_{0}
\label{ZZ}
\end{eqnarray}
where  the factor ${\rm e}^{2 \phi_4}$ has been added in order
to go from the string
to the Einstein frame~\footnote{The relation between the string and
Einstein metric is  $G_{\mu \nu}^{string} ={\rm e}^{2\phi_4}
G_{\mu \nu}^{Einstein}$.}.  $N_{\varphi}$ is a  normalization function that we
have introduced in the previous section and that will be determined by requiring
that  the super-potential is holomorphic.

The integral over the six-dimensional compact space has
been performed in Ref.~\cite{0404229} with the following result valid both
for positive and negative Chern-classes:
\begin{eqnarray}
&&\prod_{r=1}^{3}
\left[ (2 \pi R)^2 \int d^2 z_r
\sqrt{G^r} \right]   \phi_{0}^{ba}
\phi^{ab }_{0}  = \prod_{r=1}^{3} \left[
\frac{ (2 \pi R )^2  {\cal{T}}_{2}^{(r)} }{  (2 | I_r |
U_{2}^{(r)})^{1/2} } \right]  \nonumber\\
&&= (2 \pi \sqrt{\alpha'} )^6
\prod_{r=1}^{3}  \left[
\left( \frac{  T_{2}^{(r)} }{ 2 U_{2}^{(r)}} \right)^{1/2}
\left( \frac{T_{2}^{(r)} }{ | I_r | } \right)^{1/2} \right]
\label{nor54}
\end{eqnarray}
where, in going from the first to the second line, we have used Eq. (\ref{tsu})
and the last equation in (\ref{sugramo})  connecting
the four-dimensional dilaton  to the ten-dimensional one.
Inserting Eq. (\ref{nor54}) in Eq. (\ref{ZZ}) we get:
\begin{eqnarray}
Z =
\frac{{\rm e}^{  \phi_{4}}}{2} N_{\varphi}^{2} \prod_{r=1}^{3}  \left[
\left(\frac{1}{2  U_{2}^{(r)}} \right)^{1/2}
\left( \frac{T_{2}^{(r)} }{ | I_r | } \right)^{1/2} \right] =
\frac{N_{\varphi}^{2}}{2s_{2}^{1/4}}
\prod_{r=1}^{3} \left[ \frac{ 1}
{(2 u_{2}^{(r)})^{1/2}  (t_{2}^{(r)})^{1/4}} \left( \frac{T_{2}^{(r)} }{ | I_r | } \right)^{1/2} \right]
\label{Zkahler0}
\end{eqnarray}
where the last equation in (\ref{iden52}) has been used.

The scalars of the hypermultiplet can be obtained by imposing
the following conditions:
\begin{eqnarray}
\frac{|I_{1}| }{{\cal{T}}_{2}^{(1)}} =\frac{|I_{2}|}{{\cal{T}}_{2}^{(2)}}
~~~;~~I_{3} =0\,\,.
\label{n=2su}
\end{eqnarray}
When they are satisfied, it is easy to see that we have two massless excitations
corresponding to the two complex scalars of the hypermultiplet of ${\cal{N}}=2$
supersymmetry. One gets for them the following effective action:
\begin{eqnarray}
&&\! \! \! \! \! \! \! \! - \frac{1}{2g^2} \int d^4 x \sqrt{G_4} \left[
N_{\varphi_1}^{2} (D_{\mu} \varphi^{ba}_{1,-} (x) )
( D^{\mu} \varphi^{ab}_ {1,+ } (x) ) + N_{\varphi_2}^{2}
(D_{\mu} \varphi^{ba}_{2,-} (x) )
( D^{\mu} \varphi^{ab}_ {2,+} (x) ) \right]
\nonumber \\
&&\times
\prod_{r=1}^{3}
\left[ (2 \pi R)^2 \int d^2 z_r
\sqrt{G^r} \right]
\phi_{0}^{ba}
\phi^{ab }_{0 }
\label{masca7b}
\end{eqnarray}
where now the wave function
 contains only the $\Theta$-functions corresponding to the first two tori, while
the wave function along the third torus is just a constant.
{From} it, proceeding as above, we get:
\begin{eqnarray}
\prod_{r=1}^{3}
\left[ (2 \pi R)^2 \int d^2 z_r
\sqrt{G^r} \right]
\phi_{0}^{ba}
\phi^{ab }_{0 }
 = (2 \pi \sqrt{\alpha'})^6  T_{2}^{(3)}\prod_{r=1}^2
 \left[
\left( \frac{  T_{2}^{(r)} }{ 2 U_{2}^{(r)}} \right)^{1/2}
\left( \frac{T_{2}^{(r)} }{ | I_r | } \right)^{1/2} \right] .
\label{intg52}
\end{eqnarray}
Introducing, as before, the two fields:
\begin{eqnarray}
\varphi_1  = \frac{\varphi_{1,-}}{\sqrt{4\pi}}~~;~~
\varphi_2  = \frac{\varphi_{2,-}}{\sqrt{4\pi}}
\label{12}
\end{eqnarray}
we can rewrite Eq. (\ref{masca7b}) as follows:
\begin{eqnarray}
-  \int d^4 x \sqrt{G_4}  ( m,  {\bar{m}}) \left[Z^{hyper}_{1}
(D_{\mu} {\bar{\varphi}}_{1} (x) )
( D^{\mu} \varphi_ {1} (x) ) + Z^{hyper}_{2}
(D_{\mu} {\bar{\varphi}}_{2} (x) )
( D^{\mu} \varphi_ {2} (x) ) \right]
\label{masca7c}
\end{eqnarray}
where in the Einstein frame one has:
\begin{eqnarray}
Z^{hyper}_{i}  &=& \frac{{\rm e}^{2\phi_4}}{2}
{\rm e}^{- \phi_{10}} N_{i}^{2}
\prod_{r=1}^3  T_{2}^{(r)}
\prod_{r=1}^2
\frac{1 }{\left(
2  |I_r|\,\,U_{2}^{(r)}\right)^{1/2}}  \nonumber\\
&=&
\frac{N_{i}^{2}}{2  \left( 4
 u_{2}^{(1)} u_{2}^{(2)} t_{2}^{(1)}
t_{2}^{(2)} \right)^{1/2}} \,\,\, \prod_{r=1}^2
 \left(\frac{ \,T_2^{(r)}}{|I_r|}\right)^{1/2}
\label{Zn=2}\,\,.
\end{eqnarray}
The normalization factors will be determined by imposing, as in the case of chiral matter, the holomorphicity of the superpotential.

We derive also the K{\"{a}}hler metric for
the adjoint scalars.  It can be obtained by comparing Eqs. (\ref{Kaeme}) and
(\ref{meijbc}):
\begin{eqnarray}
Z_r = {\rm e}^{2 \phi_4} {\rm e}^{- \phi_{10}}
\frac{T_{2}^{(1)} T_{2}^{(2)}   T_{2}^{(3)}}{T_{2}^{(r)}U_{2}^{(r)}}=
{\rm e}^{\phi_4}
\frac{( T_{2}^{(1)} T_{2}^{(2)}   T_{2}^{(3)})^{1/2} }{
T_{2}^{(r)} U_{2}^{(r)}} = \frac{ {\rm e}^{\phi_{10}}}{ T_{2}^{(r)}
  U_{2}^{(r)} }= \frac{1}{ t_{2}^{(r)}
  U_{2}^{(r)}}
\label{Zr}
\end{eqnarray}
and this expression agrees with Eq. (2.20) of Ref.~\cite{0708.3806} obtained from the DBI action.

In the final part of this section we compute the K{\"{a}}hler metric for twisted fermions which
appears in the kinetic term for the fermions as:
\begin{eqnarray}
\frac{i}{2} \int d^4 x  Z (m , {\bar{m}} ) \sqrt{G_{4}}
\bar{\psi}^{ba} \, \gamma_{(4)}^\mu {D}_\mu \,\psi^{ab} \,\,.
\label{43b2}
\end{eqnarray}
Comparing it with Eq. (\ref{43b}) we get:
\begin{eqnarray}
Z &=&\frac{{\rm e}^{2 \phi_4}}{g^2}  N_{\psi}^{2}
\int d^6 y  \sqrt{G_{6}} (\eta^{ab})^\dag \,\eta^{ab}=
\frac{   {\rm e}^{2 \phi_4}  {\rm e}^{- \phi_{10} }}{4 \pi}  N_{\psi}^{2}
\prod_{r=1}^3
\left[\left( \frac{T_{2}^{(r)} }{ | I_r | } \right)^{1/2}
\left(\frac{T_2^{(r)}}{2  U_2^{(r)}}\right)^{1/2} \right] \nonumber\\
&=&
\frac{{\rm e}^{ \phi_4}}{4 \pi}  N_{\psi}^{2}
\prod_{r=1}^3 \left[ \left( \frac{T_{2}^{(r)} }{ | I_r | } \right)^{1/2}
\left( \frac{1}{2  U_2^{(r)} }\right)^{1/2} \right]
\label{Zfer}
\end{eqnarray}
where the factor ${\rm e}^{2 \phi_4}$  has been added for
going to the Einstein frame. Eq. (\ref{Zfer})  gives the same dependence on the
moduli as the Eq. (\ref{Zkahler0}) does, with the only difference due to a constant normalization factor.
This is an expected result which follows from
 ${\cal{N}}=1$ supersymmetry, since the fields $\varphi$ and $\psi$ belong to the same chiral multiplet.
 We also deduce that $N_{\varphi}={N_{\psi}}/{\sqrt{2\pi}}$.

In conclusion, with our procedure we have determined how the
K{\"{a}}hler metrics
explicitly  depend  on the  moduli apart from that normalization factor that
we will determine
in the next section by requiring the holomorphicity of the superpotential.

\section{Yukawa couplings}
\label{YUKA}

In this section we evaluate the Yukawa couplings both for the chiral
 multiplet and the hypermultiplet. In the case of the  chiral multiplet, we
 start from the action  in  Eq.  (\ref{LYb}) where the expansions in Eq. (\ref{exps63}) have been inserted:
\begin{eqnarray}
&&S_{3}^{(\Phi)}  =\frac{1}{2g^2}
\int d^4 x \sqrt{G_{4}}  \int d^6 y \sqrt{G_{6}}\sum_{n,m,l}
\bar{\psi}^{ca}_n\,\gamma^5_{(4)}  \nonumber \\
&& \times  \left[ \varphi_{i,\,m}^{ab}\,
\psi_l^{bc}\otimes (\eta^{ac}_n)^\dag
\gamma^i_{(6)} \phi^{ab}_{m} \eta_l^{bc} - \varphi_{i,\,m}^{bc}\,
\psi_l^{ab }\otimes (\eta^{ac}_n)^\dag
\gamma^i_{(6)} \phi^{bc}_{m} \eta_l^{ab}
\right].
\label{S31}
\end{eqnarray}
In the following we focus on the term containing the massless
scalar relative to the first torus. This
implies that the  condition:
\begin{eqnarray}
\frac{ |I_{1}^{ab} |}{ T_{2}^{(1)} }=
\frac{|I_{2}^{ab}|}{T_{2}^{(2)}} + \frac{|I_{3}^{ab}|}{T_{2}^{(3)}}
\label{sucon4}
\end{eqnarray}
must be satisfied. We are allowed to choose $I_{1}^{ab}$ being positive and consequently the massless scalar results to be  $\varphi_1=\frac{\varphi_{1,-}}{\sqrt{4\pi}}$. Furthermore, in order to satisfy the condition
\begin{eqnarray}
I_r^{ab}+I_r^{bc}+I_r^{ca}=0
\label{III}
\end{eqnarray}
and to have non-zero Yukawa couplings,
on the first torus we fix
\begin{eqnarray}
I_{1}^{ca}<0~~;~~I_{1}^{ab}>0~~;~~I^{bc}_{1}<0
\label{sceltaIa}
\end{eqnarray}
implying that the internal wave function associated with
the bosonic zero mode solution
is the first one  in Eq. (\ref{wavefu6}).

In this case, as it is shown in Appendix \ref{C},
 one is left with the following expression:
\begin{eqnarray}
S_{3}^{(\Phi)} = \int d^4 x \sqrt{G_{4}}  \bar{\psi}^{ca}\,\gamma^5_{(4)}
\varphi_{1}^{ab}\, \psi^{bc} Y^{s}
\label{YY}
\end{eqnarray}
with the Yukawa coupling in the string frame given by:
\begin{eqnarray}
Y^{s} &=& \frac{N_{\varphi_1}^{ab} N_{\psi}^{ca} N_{\psi}^{bc}}{\sqrt{2}g^2} \sqrt{4 \pi}
\prod_{r=1}^{3} \left[ (2 \pi R)^2 \int d^2 z_r
\sqrt{G^r}      \right]\nonumber\\
& \times &(\eta^{ca}_{1,-}\, \phi^{ab}_{1,+}\,
\eta^{bc}_{1,-})(\pm\eta^{ca}_{2,\mp}\,\phi^{ab}_{2,{sign I^{ab}_2}}\,
\eta^{bc}_{2,\pm})(\pm\eta^{ca}_{3,\mp}\,
\phi^{ab}_{3,{sign I^{ab}_3}}\,\eta^{bc}_{3,\pm})
\label{YY2}
\end{eqnarray}
where in each of the last  two tori we can choose the upper and lower sign
independently from each other.
The important point is that $Y^{s}$  results to be non vanishing only when the chiralities of
the two spinors $\eta^{ca}_{2,3}$ and $\eta_{2,3}^{bc}$ are opposite, while those of
$\eta_{1}^{ca}$ and $\eta_{1}^{bc}$ are equal.
As in the case of the K{\"{a}}hler metric
we add the three normalization factors for each of the three
four-dimensional fields
that we will determine by requiring the holomorphicity of the superpotential.

The integral on $T^2$
has been computed in Appendix \ref{C}. It can be generalized, by
using Eqs. (\ref{tori23}-\ref{chidef2}),
to the case of $T^2\times T^2 \times T^2$
for arbitrary values of the Chern classes as follows:
\begin{eqnarray}
 Y^{s}  &= &  \frac{{ \rm e}^{ - \phi_{10} } }{ \sqrt{8 \pi} }
\sigma{N_{\varphi}} N_{\psi} N_{\psi}
\prod_{r=1}^3 \left\{
\frac{T_{2}^{(r)} }{ \left(2 U_{2}^{(r)}|I^{ab}_r|^{\chi^{ab}_r}|I^{bc}_{r}|^{\chi^{bc}_{r}}
 |I^{ca}_{r}|^{\chi^{ca}_{r} } \right)^{1/2}}  \right.
\nonumber\\
& \times & \left.
{\Huge \Theta} \left[ \begin{array}{c} 2
\left( \frac{n'}{I^{ca}_r}+
\frac{m'}{I^{bc}_r}+\frac{l'}{I^{ab}_r}\right)
\\ 0\end{array}\right] (0|-I^{ab}_{r}
I^{bc}_{r}I^{ca}_{r} {U_f^{(r)}}) \right\}
\label{YUK}
\end{eqnarray}
with $n'=0,..., |I^{ca}|-1$; $m'=0,..., |I^{bc}|-1$; $l'=0,..., |I^{ab}|-1$. Moreover
$\chi_r$ is defined in Eq.(\ref{chidef}) and
\begin{eqnarray}
U_f^{(r)}=\left\{\begin{array}{c} U^{(r)}\quad{\rm for} \quad sign(I^{ca}I^{bc}I^{ab})<0\\
{\bar U}^{(r)}\quad {\rm for}\quad sign(I^{ca}I^{bc}I^{ab})>0\,\,.\end{array}\right. \label{56}
\end{eqnarray}
The previous result agrees with the one in Ref.~\cite{0404229} and each of the  three $\Theta$-functions in Eq. (\ref{YUK}) is a holomorphic function of the complex structure of the corresponding torus. It is worth noticing that, because of Eq. (\ref{56}), if $sign (I^{ca}_r I^{bc}_r I^{ab}_r )$
is not the same for each value of $r=1,2,3$, then some of $\Theta$-functions
will depend on $U$, while others will depend on ${\bar{U}}$.

Notice that Eq. (\ref{YUK}) is valid not only for the choice in Eq. (\ref{sceltaIa}), but for any arbitrary choice
of the Chern classes.
One can rewrite the Yukawa couplings in the Einstein frame
multiplying the  equation by
 ${\rm e}^{4  \phi_4}  {\rm e}^{-\phi_4}$,
where the first factor comes from the rescaling of the square root of the
determinant of the metric and the second from that of the two fermionic fields.
Using the last equation in  (\ref{sugramo}) and
taking then into account  Eq. (\ref{expK2}), we see that the factors
containing $\phi_4$ combine together with other factors to give:
\begin{eqnarray}
&&Y^{E}  =\frac{{\rm e }^{K/2}}{\sqrt{8\pi}}\sigma
{N_{\varphi}} N_{\psi} N_{\psi}
\prod_{r=1}^3  \left\{ \frac{(T_{2}^{(r)})^{1/2} }{
\left(2
|I^{ab}_r|^{\chi^{ab}_r}|I^{bc}_{r}|^{\chi^{bc}_{r}}
 |I^{ca}_{r}|^{\chi^{ca}_{r} } \right)^{1/2}}
\right. \nonumber \\
&& \times \left.
{\Huge \Theta} \left[ \begin{array}{c}
2\left(\frac{n'}{I^{ca}_r}+
\frac{m'}{I^{bc}_r}+\frac{l'}{I^{ab}_r}\right)
\\ 0\end{array}\right] (0|-I^{ab}_{r}
I^{bc}_{r}I^{ca}_{r} {U_f^{(r)}}) \right\}
\label{YUKEINa}
\end{eqnarray}
where $K$ is the K{\"{a}}hler potential given in Eq. (\ref{KAE}).
With the choice in Eqs. ({\ref{sucon4}}) and ({\ref{sceltaIa}) the previous equation becomes:
\begin{eqnarray}
Y^{E} &=&\frac{{\rm e }^{K/2}}{\sqrt{8\pi}}\sigma
{N^{ab}_{\varphi_1}} N_{\psi}^{ca} N_{\psi}^{bc}
\frac{ (T_{2}^{(1)})^{1/2} }{\left(2
I^{ab}_1\right)^{1/2}} \,\,\,
\prod_{r=2}^3
\frac{(T_{2}^{(r)})^{1/2} }{ \left(2
|I^{bc}_{r}|^{\chi^{bc}_{r}}
 |I^{ca}_{r}|^{\chi^{ca}_{r} } \right)^{1/2}}
\nonumber \\
&\times&
\prod_{r=1}^3  {\Huge \Theta} \left[ \begin{array}{c}
2\left(\frac{n'}{I^{ca}_r}+
\frac{m'}{I^{bc}_r}+\frac{l'}{I^{ab}_r}\right)
\\ 0\end{array}\right] (0|-I^{ab}_{r}
I^{bc}_{r}I^{ca}_{r} {U_f^{(r)}}) \,\,.
\label{YUKEIN}
\end{eqnarray}
Because of the terms depending on  the magnetizations, Eq. (\ref{YUKEIN})  is not a holomorphic function of the moduli unless we choose the
normalization factors $N_{\varphi_1}^{ab}, N_{\psi}^{bc}, N_{\psi}^{ca}$
in such a way to eliminate such dependence.
This is what we are going to explain in the following.

Eq. (\ref{III}) must be satisfied for the three tori. In the first torus we have
chosen the $I$'s as in Eq. (\ref{sceltaIa}). In the second torus let us choose
$sign (I^{ab}_{2})={ sign}(I^{bc}_{2})$~\footnote{If this is not
the case, we can repeat what we are going to do, substituting
$I_{2}^{bc}$ with $I_{2}^{ca}$ without loss of generality.}.
With these two choices  we get:
\begin{eqnarray}
&&I^{ab}_{1}+I^{bc}_{1}+I^{ca}_{1}=0\Longrightarrow |I^{bc}_{1}|+
|I^{ca}_{1}|=|I^{ab}_{1}|
\Longrightarrow  \nu^{ab}_{1} =\nu^{bc}_{1} +\nu^{ca}_{1}\label{T1}\\
&&I^{ab}_{2}+I^{bc}_{2}+I^{ca}_{2}=0\Longrightarrow |I^{ab}_{2}|+
|I^{bc}_{2}|=|I^{ca}_{2}|
\Longrightarrow  \nu^{ca}_{2}=\nu^{ab}_{2} +\nu^{bc}_{2}\label{T2aa}
\end{eqnarray}
because $sign(I^{ca}_2)=-sign(I^{bc}_2)$, as follows from
Eq. (\ref{YY2}). In the last step we have used the definition:
\begin{eqnarray}
\pi\nu_r \equiv \frac{|I_r|}{T_2^{(r)}}
\label{pinu}
\end{eqnarray}
where the quantities $\nu_r$ will be shown to have also a precise meaning in the theory of magnetized branes and strings generating our model at low-energy.
Finally,  on the third torus let us choose $sign (I^{ab}_{3})=-{ sign}(I^{bc}_{3})$.
This means that:
\begin{eqnarray}
I^{ab}_{3}+I^{bc}_{3}+I^{ca}_{3}=0\Longrightarrow |I^{ab}_{3}|+
|I^{ca}_{3}|=|I^{bc}_{3}| \Longrightarrow  \nu^{bc}_{3} =
\nu^{ab}_{3} +\nu^{ca}_{3}
\label{T3}
\end{eqnarray}
because $sign(I^{ca}_3)=-sign(I^{bc}_3)$, as follows from
Eq. (\ref{YY2})~\footnote{For any other choice of signs we get either an
equivalent  realization of ${\cal{N}}=1$ supersymmetry
or an extended ${\cal{N}}=4$
supersymmetry that we do not consider here.}.

Summing Eqs.  (\ref{T1}), (\ref{T2aa}) and  (\ref{T3})  we get:
\begin{eqnarray}
(\nu^{ab}_{1} -\nu^{ab}_{2}-\nu^{ab}_{3}) -
( \nu^{bc}_{1}+\nu^{bc}_{2} -\nu^{bc}_{3})+
(-\nu^{ca}_{1}+\nu^{ca}_{2}-\nu^{ca}_{3})=0
\label{summing}
\end{eqnarray}
which is satisfied by taking:
\begin{eqnarray}
\nu^{ab}_{1} =\nu^{ab}_{2}+\nu^{ab}_{3}~~;~~
\nu^{bc}_{3}=\nu^{bc}_{1} +\nu^{bc}_{2}~~;~~
\nu^{ca}_{2}= \nu^{ca}_{1}+\nu^{ca}_{3}\,\,.
\label{presu1}
\end{eqnarray}
Such a configuration preserves ${\cal N}=1$ supersymmetry
 in all the three  sectors
$ab$, $bc$ and $ca$.  With the previous choices one gets:
\begin{eqnarray}
\chi_{2}^{bc} = \chi_{3}^{ca} =0~~;~~\chi_{3}^{bc} = \chi_{2}^{ca} =1 \,\, .
\label{chis}
\end{eqnarray}
Using these values in Eq. (\ref{YUKEIN}) we see that, if the normalization factors are taken  as follows:
\begin{eqnarray}
N_{\varphi_1}^{ab}=\left(\frac{|I^{ab}_{1}|}{T_2^{(1)}}\right)^{1/2}~~;~~
N_{\psi}^{ca}=\left(\frac{|I^{ca}_{2}|}{T_2^{(2)}}\right)^{1/2}~~;~~
N_{\psi}^{bc}= \left(\frac{|I^{bc}_{3}|}{T_2^{(3)}}\right)^{1/2}
\label{scelta}
\end{eqnarray}
then the Yukawa coupling becomes a holomorphic function of the moduli!

In the final part of this section we consider the Yukawa coupling involving the
two fermions of the hypermultiplet and a scalar living in the adjoint
representation of the gauge
group~\footnote{We call it adjoint with an abuse of notation having in mind that
the $U(1)$ gauge group is extended to a non-abelian group when some of the background values are equal to each other as discussed in Sect. \ref{2}.}.
This coupling is  obtained by compactifying the terms of ten dimensional action
given in Eq. (\ref{LYbc}).  In the following we restrict our analysis
only to the first term of this equation which gives the interaction  of the two fermions $\bar{\psi}^{ab}_\alpha$ and ${\psi}_\beta^{ba}$
living in the bifundamental representation  of the gauge group
$U_a(1)\times U_b(1)$  with the massless scalar in the
``adjoint" representation of the second
gauge group. Here, the indices $\alpha$ and $\beta$ label the
degeneracy of the lowest fermionic state as described at the end
of Appendix \ref{B}.
The second term of  Eq. (\ref{LYbc}) gives the  interaction of the two
fermions with the scalar in the ``adjoint" of the group $U_a (1)$. Such
interaction term has a sign that is   opposite to the first term. This sign can be
taken into  account by multiplying the Yukawa coupling by a factor
$\sigma$ which is equal to $+1\,\,[-1]$ if the fermion $\psi^{ba}$ is in the
fundamental representation of the gauge group $U_{b}(1)\,\, [U_a (1)]$.

Inserting in the first term of Eq.  (\ref{LYbc})  the zero mode  of  the
expansion in   Eq. (\ref{kkre4}) and the massless fermions given in
Eq. (\ref{wavef})
we get:
\begin{eqnarray}
{\cal S}^{\delta B}_{3;\,\alpha,\,\beta}&=&  \int d^4 x \sqrt{G_4}
\bar{\psi}^{ab}_\alpha \, \sum_{r=1}^3 [\varphi_r^a ~ (Y^{r}_{\alpha,\,\beta})^s
+ \bar{\varphi}_r^a~(\bar{Y}^{r}_{\alpha,\,\beta})^s ]~
\gamma^5_{(4)} \psi^{ba}_\beta
\label{n=2yuk}
\end{eqnarray}
where we have taken the internal wave function of the scalar to be equal to 1
and used the relation between the four-dimensional fields and ten-dimensional ones given in Eq. (\ref{czvarphi}). The Yukawa coupling in the four-dimensional
string frame  is equal to:
\begin{eqnarray}
(Y_{\alpha,\,\beta}^{r})^s=
\sqrt{4\pi}\,\sigma \frac{\sqrt{\alpha'}}{R}\, \frac{N_{\psi_\alpha} N_{\psi_\beta}}{2g^2}\prod_{r=1}^3 \left[ (2\pi R)^2\int d^2 z^r \sqrt{G^{(z^r,\,\bar{z}^r)}} \right](\eta^{ba}_\alpha)^\dag \left[ \frac{1}{2U_2^{(r)}} \gamma^{z^r}_{(6)}\right] \eta^{ba}_\beta
\label{Yrab}
\end{eqnarray}
with $(\bar{Y}^{r}_{\alpha,\,\beta})^s = ((Y_{\beta,\,\alpha}^{r})^\dag)^s$.
The  factors depending on $R$ and $\sqrt{\alpha'}$  are a direct
consequence of the factors present in Eq. (\ref{czvarphi}).

Due to the peculiar structure of the six-dimensional $\Gamma$ matrices, the only
terms of the previous expression, that are   different from zero, are the ones
with $r=3$  and $\alpha\neq\beta$. The result is:
\begin{eqnarray}
(Y_{\uparrow,\,\downarrow}^{3})^s=
(Y_{\downarrow,\,\uparrow}^{3})^s=
\sigma\,N_{\psi_\downarrow} N_{\psi_\uparrow}\frac{ e^{-\phi_{10}} }{2\sqrt{4\pi}}\,
\prod_{r=1}^3 T^{(r)}_2  \prod_{r=1}^2 \left[\frac{1}{(2U_2^{(r)}
|I_r^{ab}|)^{1/2}}\right] \sqrt{\frac{1}{U_2^{(3)}{ T}^{(3)}_2}}
\label{YUKH}
\end{eqnarray}
where we have used the relation between $T_2$ and ${\cal T}_2$ given in Eq. (\ref{tsu}) together with the expression of the Dirac-matrix given in Eq. (\ref{curvega}).

The Yukawa coupling in the Einstein frame is obtained by multiplying
the previous  expression by  the dilaton factor $e^{3\phi_4}$.
Introducing, as for the case ${\cal N}=1$, the K\"ahler potential $K$ given in Eq.
(\ref{expK2}) we get:
\begin{eqnarray}
(Y_{\uparrow,\,\downarrow}^{3})^E =(Y_{\downarrow,\,\uparrow}^{3})^E=
\sigma\,N_{\psi_\downarrow} N_{\psi_\uparrow} \frac{ e^{K/2} }{4 \sqrt{4\pi} }\,
\prod_{r=1}^2 \left( \frac{T_2^{(r)}}{ |I_r^{ab} |}\right)^{1/2}\,\,.
\label{Yupdown}
\end{eqnarray}
The previous coupling is not a holomorphic function of the moduli.
However, normalizing the fermions as follows:
\begin{eqnarray}
N_{\psi_\uparrow}=N_{\psi_\downarrow}= \left(\frac{|I^{ab}_1|}{T_2^{(1)}}\right)^{1/2}=\left(\frac{|I^{ab}_2|}{T_2^{(2)}}\right)^{1/2}
\label{NpsiN}
\end{eqnarray}
where the first relation in Eq.  (\ref{n=2su}) has been explicitly used, we restore
also in this case the holomorphicity of the super-potential.

\section{Fixing the K{\"{a}}hler metrics}
\label{FIX}

In the previous section we have fixed the normalization of the
four-dimensional
fields by requiring that the Yukawa couplings come from a
holomorphic superpotential. We can now go back to the K{\"{a}}hler metrics
that we have determined in Sect. \ref{Kaehler}  apart from an overall normalization,
and fix them uniquely using the value  obtained from  the Yukawa couplings. Let us start from the chiral multiplet where the    K{\"{a}}hler metric
is given in Eq. (\ref{Zkahler0}). Inserting in it the normalization given in the first
equation in (\ref{scelta}) we get:
\begin{eqnarray}
Z^{chiral}_{ab}  =
\frac{1}{2 s_{2}^{1/4}}
\prod_{r=1}^{3} \left[ \frac{ 1}
{(2 u_{2}^{(r)})^{1/2}  (t_{2}^{(r)})^{1/4}} \right]
\left( \frac{\nu_{1}^{ab}  }{\pi  \nu_{2}^{ab} \nu_{3}^{ab} } \right)^{1/2}
\label{Zkaehlerf}
\end{eqnarray}
where Eq. (\ref{pinu}) has been used. Let us now examine the dependence
on the magnetizations, given by the last factor in the r.h.s.
of the previous equation. The dependence of the K{\"{a}}hler metric on the
magnetizations has been computed by means of a pure string calculation in
Refs.~\cite{0404134,0512067} obtaining in  our notations:
\begin{eqnarray}
\left[ \frac{\Gamma(1-\nu^{ab}_{1})}{\Gamma(\nu^{ab}_{1})}
\frac{\Gamma(\nu^{ab}_{2})}{\Gamma(1 - \nu^{ab}_{2})}
\frac{\Gamma(\nu^{ab}_{3})}{ \Gamma(1 - \nu^{ab}_{3})} \right]^{1/2}
\Longrightarrow  \left( \frac{\nu_{1}^{ab}  }{
\nu_{2}^{ab} \nu_{3}^{ab} } \right)^{1/2}\,\,.
\label{gam6}
\end{eqnarray}
This expression, in the limit of small magnetizations, coincides with our
result in Eq. (\ref{Zkaehlerf}) for the part concerning the magnetizations, consistently
with the fact that this is just the limit that
one should perform in going from the string to the field theory
definition of $\nu$.  In this limit, for positive values of $\nu$, one has:
\begin{eqnarray}
\tan \pi \nu_r = \frac{|I_r|}{T_{2}^{(r)}} \Longrightarrow  \pi \nu_r =
\frac{|I_r|}{T_{2}^{(r)}}
\label{ftlim}
\end{eqnarray}
which is realized for small values of $\nu_r$.  The expression for the twisted
K{\"{a}}hler metric obtained from considerations about holomorphicity
within the instanton calculus~\cite{0705.2366,0709.0245,0711.0866}
contained a possible additional dependence
on the magnetizations that we do not find in our field theoretical procedure.

Turning to the K{\"{a}}hler metric of the hypermultiplet, given in Eq. (\ref{Zn=2}),
we see that the dependence on the magnetization cancels and we get:
\begin{eqnarray}
Z^{hyper}_{i}  =
\frac{1}{2 \left( 4
 u_{2}^{(1)} u_{2}^{(2)} t_{2}^{(1)}
t_{2}^{(2)} \right)^{1/2}} \,\,.
\label{Zn=2f}
\end{eqnarray}
In this case the dependence on the magnetization drops out in agreement
with the well-known result (see for instance Eq. (2.45) of
Ref.~\cite{0708.3806}).

\section{Conclusions and outlook}
\label{conclu}

In this paper we have proposed a procedure for determining the K{\"{a}}hler
metric for the twisted open strings defined as the ones having their end-points attached to two
D branes with different magnetizations. Unlike Ref.~\cite{0404229},  where
the kinetic terms are canonically normalized and then the K{\"{a}}hler metrics
appear in the Yukawa couplings, we keep for the quadratic terms the normalization
that comes naturally from the Kaluza-Klein reduction apart from a
normalization factor that we then determine requiring that the Yukawa couplings
correspond to a holomorphic superpotential. We find that these
normalization factors depend only on the magnetization.  This procedure
yields the K{\"{a}}hler metrics proposed in the
literature~\cite{0705.2366,0709.0245,0711.0866} without the arbitrary factors
that appeared in the previously mentioned  proposals. In particular, our procedure
allows us to correctly determine the K{\"{a}}hler metric for the hypermultiplet
that agrees with the expression obtained  with other methods.

In deriving the previous results we
have, however, made implicitly two assumptions. The first one is that
the normalization
factor contains only the minimal number of factors that make the superpotential
holomorphic and the second one is that our reasonings are based
on  the specific form of the scalar fields $\varphi_{r, \pm}$  that we use
(see Eqs. (\ref{varphi+-}) and  (\ref{+-87})).
But why do we use these scalar fields? Before trying to answer  this question,
let us observe that the  introduction of  the normalization factor  allows us
to actually rescale the field with a quantity and at the same time rescale the
normalization factor with the inverse quantity without changing the K{\"{a}}hler
metrics and the Yukawa couplings.  In particular, this rescaling factor
can be a function of the moduli.  This means that the presence of
the normalization factor does not allow us
to determine the absolute normalization
of the scalar field.

Having said this, let us find  the relation of  $\varphi_{r,-}$  with the original
ten-dimensional fields.  In the case of the adjoint scalars such relation
is given  in Eq. (\ref{varphir}). For the twisted fields, starting from Eq.
(\ref{varphi+-}) and then using the relation between
the variables $x^1 , x^2$  and $z, {\bar{z}}$
given in Appendix  \ref{T2},  we get:
\begin{eqnarray}
\varphi_{r\,-} &=&\sqrt{ \frac{ 2  \,U_2^{(r)} }{{\cal T}_2^{(r)}}}
\varphi_{r\,z}
=\sqrt{ \frac{ 2  \,U_2^{(r)} }{{\cal T}_2^{(r)}}} \left(\frac{\partial
x^{2r+2} }{2\pi R \,\partial z^r}
W_{2r+2}+  \frac{\partial x^{2r+3} }{2\pi R\,\partial z^r}
W_{2r+3}\right) \nonumber\\
&=& \frac{i}{\sqrt{2U_2^{(r)}{\cal T}_2^{(r)}} }(\bar{U}^{(r)}
W_{2r+2}-W_{2r+3})
\label{zphi}
\end{eqnarray}
where we have used Eqs.   (\ref{comcor}) and the
transformation rule  of a covariant vector~\footnote{The extra factor
$2\pi R$ in Eq. (\ref{zphi})  is necessary  for dimensional
reasons (the  x variables  are dimensional, while the z variables
are dimensionless).}:
\begin{eqnarray}
W_{z^r} = \frac{\partial x^k}{2\pi R\partial z^r }W_{k} \,\,.
\label{Wzr}
\end{eqnarray}
Unlike the adjoint scalar in Eq. (\ref{varphir}), the fundamental scalar
in Eq. (\ref{zphi}) is not a holomorphic function of the fields for the presence
of the non-holomorphic pre-factor.  If we want a holomorphic function we
can incorporate  the extra non-holomorphic factor in the normalization factor
fixing it in a unique way. This requirement
 eliminates the possibility of rescaling both the scalar field and the normalization
factor, as discussed above. This unique rescaling leaves
both the K{\"{a}}hler metric and the Yukawa couplings, determined above,
unchanged.

The procedure outlined in this paper can be extended to more complicated
and more realistic compact manifolds and, if we restrict ourselves to toroidal
compactifications, it would be important to develop string techniques for
fully reproducing the field theoretical results in the zero slope limit and also
for computing string corrections to the field theory behavior.

\vspace{.5cm}

\begin{center}
{\bf Acknowledgments}
\end{center}
\noindent
We thank M. Berg, M. Bill{\'{o}}, E. Brynjolfsson, D. L\"{u}st, M. Frau, A. Lerda, I. Pesando,
 R. Russo, L. Thorlacius and F. Zwirner for discussions. R.M. and F.P. thank the Niels Bohr Institute and Nordita for their kind hospitality.

\appendix

\section{The torus $T^2$}
\label{T2}

In this Appendix we summarize the properties of the torus $T^2$ and we list the
combination of the string moduli that enter in supergravity.

The torus $T^2$ can be equivalently described either by the  ``curved"
dimensional coordinates $x^1 , x^2$ that are periodic with period
$2\pi R$ going around the
two one-cycles of the torus
\begin{eqnarray}
x^1\equiv x^1+ 2 \pi R \qquad x^2\equiv x^2+ 2 \pi  R
\label{lattid}
\end{eqnarray}
or by  the ``flat"  dimensionless coordinates $z, {\bar{z}}$  given by:
 \begin{eqnarray}
z= \frac{x^1 + U x^2}{ 2 \pi R}   \qquad
\bar{z}= \frac{ x^1 +  \bar{U} x^2}{2 \pi  R} \,\,.
\label{comcor}
\end{eqnarray}
The dimensional parameter $R$ is arbitrary and has been introduced to  deal
with dimensionless  $z$ and ${\bar{z}}$. We will see, however, that the physical
quantities do not depend on $R$.
The metric of the torus in the two coordinate systems is equal to:
 \begin{eqnarray}
G_{ij}^{(x^1 , x^2 )} = \frac{{\cal{ T}}_2}{U_2} \left(   \begin{array}{cc} 1 & U_1 \\
                U_1 & |U|^2 \end{array} \right)~~;~~
                G_{ij}^{(z  , {\bar{z}})} = \frac{{ \cal{T}}_2}{2 U_2}
                \left(   \begin{array}{cc} 0 & 1 \\
                1 & 0 \end{array} \right) \,\, .
\label{meGG}
\end{eqnarray}
They imply
\begin{eqnarray}
ds^2 = G_{ij}^{(x^1 , x^2 )} d {x^i}  d x^j
=  \frac{ {\cal{T}}_2}{U_2}  | d {x^1}  + U d {x^2} |^2 =
(2 \pi R)^2 \frac{{ \cal{T}}_2}{U_2}  dz d {\bar{z}} \,\,.
\label{ds2}
\end{eqnarray}
The complex quantities $U = U_1 +i U_2$ and ${\cal{T}} =
{ \cal{T}}_1 + i { \cal{T}}_2$
correspond, respectively, to the complex
and the K{\"{a}}hler structures of the torus $T^2$. The real part of the K{\"{a}}hler structure ${\cal{T}}_1$ is related to the the Kalb-Ramond field by
${\cal{T}}_1 = - B_{12}$, while its imaginary part is related to the volume
of the torus.  We use
dimensionless moduli. They are given in terms of the physical parameters
of the torus, consisting of two radii $R_1$ and $R_2$ and an angle
$\alpha$, by the following expressions:
\begin{eqnarray}
U = \frac{R_2}{R_1} {\rm e}^{i \alpha }~~;~~{\cal{T}}_2  =
\frac{R_1 R_2}{R^2} \sin \alpha \,\,.
\label{UT2}
\end{eqnarray}
The area  of the torus $T^2$ is given by:
\begin{eqnarray}
&&{\cal{A}} = \int_{0}^{2 \pi R} d x^1 \int_{0}^{2 \pi R} d x^2 \sqrt{G^{(x^1 , x^2)}} =
(2 \pi)^2 R_1 R_2 \sin \alpha  \nonumber\\
&&=  (2 \pi R)^2 \int_{T^2} d^2 z  \sqrt{G^{(z , {\bar{z}})}} = (2 \pi R)^2 {\cal{T}}_2
\label{Area9}
\end{eqnarray}
and is independent of $R$.

In string theory one usually introduces ``curved'' dimensional
coordinates $\tilde{x}^1$, $\tilde{x}^2$  which have periodicity
$2\pi \sqrt{\alpha'}$ when translated along the two one-cycles
of the torus~\footnote{For the sake of simplicity we could have introduced torus coordinates with the same periodicity $R = \sqrt{\alpha'}$. It is, however, useful
to keep them different from each other to have a check on the formulas because
the physics is independent on their choice.}.
The relation between these coordinates and the
ones given in Eq. (\ref{lattid}) is $\tilde{x}=(\sqrt{\alpha'}/R) x$. In terms of these coordinates the volume of the torus is measured in units of the string length $2\pi\sqrt{\alpha'}$ rather then $2\pi R$.
This means that the
string K\"ahler structure $T_{2}$  is given by Eq. (\ref{UT2})
with $R$ substituted by $\sqrt{\alpha'}$, i.e.
\begin{eqnarray}
 T_{2}  =   \frac{R^2}{\alpha'} {\cal{T}}_{2} \,\, .
\label{tsu}
\end{eqnarray}
The relation between the covariant fields defined in the $x$-coordinates with the corresponding ones  in the $\tilde{x}$-coordinates can be obtained from the general transformation of coordinates rule of covariant fields:
\begin{eqnarray}
\tilde{C}_r=\left(\frac{\partial x^s}{\partial\tilde{x}^r}\right)
C_s=\frac{{R}}{\sqrt{\alpha'}} C_r \,\,.
\label{tildeCr}
\end{eqnarray}
It is also useful to give the relation between the four-dimensional  field $\varphi$ defined in Eq. (\ref{varphir}) and the scalars written in the complex system of coordinates $C_{z}$. From its definition, given in Eq. (\ref{varphir}), we can write:
\begin{eqnarray}
\varphi= i\frac{R\bar{U}}{\,\sqrt{4\pi\alpha'}} \left(\frac{2\pi R\partial z }{\partial {x}^{1}}\right)
C_z-i \frac{R}{\sqrt{4\pi\alpha'}} \left(\frac{2\pi R\partial z }{\partial {x}^{2}}\right)C_z
= \frac{2U_2R}{\sqrt{4\pi\alpha'}}C_z
\label{czvarphi}
\end{eqnarray}
where the extra factor $2\pi\,R$ is necessary for dimensional reasons, because the $z$'s, differently from the $x$'s, are dimensionless coordinates.

We can introduce the following vierbein and its inverse:
\begin{eqnarray}
e^I_{~i}=\frac{1}{2}\sqrt{\frac{{\cal T}_2}{U_2}}\left( \begin{array}{cc}
                                 1 &1\\
                                 -i&i
                                 \end{array}\right)~~;~~i=z,\,\bar{z}~~;~~
                                 e^i_{~I}= \sqrt{\frac{U_2}{{\cal T}_2}}\left( \begin{array}{cc}
                                 1 & i\\
                                 1 &-i
                                 \end{array}\right)
                                 \label{vierbe2}
\end{eqnarray}
such that
\begin{eqnarray}
G^{(z, {\bar{z}})}_{ij}=e^J_{~i}\delta_{JI}e^I_{~j}
\equiv (e^t\,e)_{ij}=\frac{{\cal T}_2}{2U_2}\left( \begin{array}{cc}
                                 0 &1\\
                                 1&0
                                 \end{array}\right)\,\,.
                                 \label{Gfrome}
\end{eqnarray}
In the final part of this Appendix we introduce the moduli fields that one should use
in the supergravity action. In string theory the moduli are the
ten dimensional dilaton and the ones related to the complex and K{\"{a}}hler structure
$U= U_1 + i U_2$ and $T = T_1 + i T_2$. In supergravity the variables to use are
instead the following:
\begin{eqnarray}
s_2 = {\rm e}^{- \phi_{10}} \prod_{r=1}^{3} T_{2}^{(r)}~~;~~
t_{2}^{(r)} = {\rm e}^{- \phi_{10}} T_{2}^{(r)}~~;~~u_2 = U_2~~;~~
{\rm e}^{- \phi_{4}} = {\rm e}^{- \phi_{10}} \prod_{r=1}^{3}
\left(T_{2}^{(r)} \right)^{1/2}
\label{sugramo}
\end{eqnarray}
The subindex 2 means that they are the imaginary part of a complex
quantity whose real part is given by suitable RR fields that are not needed here.
The previous relations imply the following:
\begin{eqnarray}
\frac{s_2}{ \prod_{r=1}^{3} t_{2}^{(r)}} = {\rm e}^{2  \phi_{10}}~~;~~
 \prod_{r=1}^{3} T_{2}^{(r)} = \frac{s_{2}^{3/2} }{ (\prod_{r=1}^{3} t_{2}^{(r)})^{1/2}}
 ~~;~~{\rm e}^{2 \phi_4}  = s_{2}^{-1/2} (\prod_{i=1}^{3} t_{2}^{(i)})^{-1/2} \,\, .
\label{iden52}
\end{eqnarray}
The K{\"{a}}hler potential of the closed string moduli is given by:
\begin{eqnarray}
K = - \log s_2 - \sum_{r=1}^{3} \left[ \log t_{2}^{(r)}  + \log  u_{2}^{(r)}   \right] \,\,.
\label{KAE}
\end{eqnarray}
It satisfies the following identity:
\begin{eqnarray}
e^{K/2}=\frac{e^{2\phi_4}}{\prod_{r=1}^3(U_2^r)^{1/2}} \,\,.
\label{expK2}
\end{eqnarray}

\section{Solving the eigenvalue equations}
\label{B}

Let us start analyzing the case of the torus $T^2$.
In terms of the variables $z, {\bar{z}}$ defined in the Appendix \ref{T2}  the
gauge covariant derivative is given by:
\begin{eqnarray}
{\tilde{D}}_{z} = \partial_{z} - i B_z~~;~~{\tilde{D}}_{{\bar{z}}} =
\partial_{{\bar{z}}}  - i  B_{\bar{z}}
\label{DzDzbar}
\end{eqnarray}
where the background fields $B_z$ and $B_{{\bar{z}}}$ are given by:
\begin{eqnarray}
B_z = \frac{\pi  I {\bar{z}}}{ (U - {\bar{U}})}~~;~~
B_{\bar{z}} = - \frac{\pi  I {{z}}}{ (U - {\bar{U}})}\Longrightarrow
B = B_z dz + B_{\bar{z}} d{\bar{z}}= \frac{ \pi I
( {\bar{z}} dz - z d {\bar{z}})}{ 2i U_2} \,\,.
\label{AzAzbar}
\end{eqnarray}
They imply $(F \equiv dB)$:
 \begin{eqnarray}
 \left[  -i {\tilde{D}}_{{{z}}} ,   -i  {\tilde{D}}_{{\bar{z}}} \right] =
 - \frac{\pi I}{U_2} \equiv i F_{z {\bar{z}}}\,\,.
 \label{commu9I}
\end{eqnarray}
The expression for $F_{z {\bar{z}}}$ can be obtained from the fact that the first Chern class must be an integer $I$:
\begin{eqnarray}
\int \frac{F}{2 \pi} = \int F_{z {\bar{z}} } dz \wedge d{\bar{z}} =I
\Longrightarrow F_{z{\bar{z}}} = - \frac{\pi I}{i U_2}\,\,.
\label{Chern}
\end{eqnarray}
{From} the previous expression for $ F_{z{\bar{z}}}$ one can easily compute
\begin{eqnarray}
2i<(F_r)^{ {{z}}}_{\,\,{ z}} >^{ab} = 2{i}G^{(r) z {\bar{z}}}
<(F_r)_{ {\bar{z}} { z}} >^{ab} = - \frac{4i U_{2}^{(r)}}{{\cal{T}}_{2}^{(r)}}
(F_r)_{z {\bar{z}}}^{ab}
=  \frac{4 \pi I_r}{{\cal{T}}_{2}^{(r)}}
\label{Gzz63}
\end{eqnarray}
and
\begin{eqnarray}
2i<(F_r)^{ {\bar{z}} }_{\,\,\bar{ z}} >^{ab} =2i
G^{(r) {\bar{z}} z}   <(F_{r})_{z  {\bar{z}}}>^{ab}  =
\frac{4i U_{2}^{(r)}}{{\cal{T}}_{2}^{(r)} }(F_{r}^{ab})_{z {\bar{z}}} =
- \frac{4 \pi I_r}{{\cal{T}}_{2}^{(r)}}
\label{GG564}
\end{eqnarray}
where  the metric $G^r$  given in Eq. (\ref{meGG}).

We also introduce
\begin{eqnarray}
I^{ab}=I^a-I^b=-i\frac{U_2}{\pi}(F^{a}_{z {\bar{z}}}-F^{b}_{z {\bar{z}}}) \,\,.
\label{Iab}
\end{eqnarray}
If one considers  the quadratic terms in the action, there is no loss of generality in choosing the magnetization
on the $D$ brane labeled  with the index $b$ to be zero. This allows us
to simplify the notation by writing $I^{ab}=I^a\equiv I$.
We perform  this choice in Sects. \ref{2}  and
 \ref{Kaehler} while we will reintroduce the indexes   when considering the
Yukawa couplings in Sect. \ref{YUKA}.

Using the metric for the torus $T^2$ given in Appendix \ref{T2} one
gets~\footnote{In this equation and in the entire analysis of the torus $T^2$
the index $k$ runs only on one torus and should not be confused with the one
used in Eq. (\ref{eq88}).} :
\begin{eqnarray}
 {\tilde{D}}_{k}  {\tilde{D}}^{k} = {\tilde{D}}_{k}  G^{ki} {\tilde{D}}_{i} =
\left( \begin{array}{cc}  {\tilde{D}}_{{{z}}} & {\tilde{D}}_{{\bar{z}}} \end{array}
\right)  \frac{2 U_2}{{\cal{T}}_2} \left( \begin{array}{cc} 0 & 1 \\
         1 & 0 \end{array} \right) \left( \begin{array}{c}  {\tilde{D}}_{{{z}}} \\
         {\tilde{D}}_{{\bar{z}}} \end{array} \right)  =    \frac{2 U_2}{{\cal{T}}_2}
         \left\{  {\tilde{D}}_{{{z}}} ,    {\tilde{D}}_{{\bar{z}}} \right\} \,\,.
\label{Delta2}
\end{eqnarray}
If $I >0$ we can introduce the creation and annihilation operator:
\begin{eqnarray}
 -i {\tilde{D}}_{{{z}}} \equiv -i \left( \partial_z - \frac{\pi I {\bar{z}}}{2U_2 }\right)
 = \sqrt{\frac{\pi I}{U_2}} a^{\dagger}~~;~~
 -i  {\tilde{D}}_{{\bar{z}}} \equiv -i \left(  \partial_{\bar{z}} +
 \frac{\pi I {{z}}}{2U_2 } \right)   =\sqrt{\frac{\pi I}{U_2}} a
\label{aadagger}
\end{eqnarray}
that  satisfy the harmonic oscillator algebra:
\begin{eqnarray}
[ a , a^{\dagger} ] =1\,\,.
\label{aadaggerb}
\end{eqnarray}
Using Eqs. (\ref{aadagger}) in Eq. (\ref{Delta2}) we get:
\begin{eqnarray}
-{\tilde{D}}_{k}  {\tilde{D}}^{k} =  \frac{2 \pi I}{{\cal{T}}_2} \left(
a a^{\dagger} + a^{\dagger} a \right) =
 \frac{2 \pi I}{{\cal{T}}_2}  \left(  2 a^{\dagger} a +1 \right) \equiv
 \frac{2 \pi I}{{\cal{T}}_2} \left( 2N +1 \right) \,\,.
\label{DD96}
\end{eqnarray}
The ground state for the torus $T^2$  is degenerate
and there are $I$ independent solutions given by:
\begin{eqnarray}
\phi^{ab,n}_{T^2} (z) =  e^{\pi iI {{z}}  \frac{ {\rm Im}z}{ {\rm Im} U  }}\,
\Theta\left[\begin{array}{cc} \frac{2n }{ I}\\0\end{array}\right]
(I  z  | I  U ) ~~;~~n  =0\dots I  -1
\label{wafu5}
\end{eqnarray}
which are determined by solving the equation
\begin{eqnarray}
a \,\, \phi^{ab}_{T^2} (z, {\bar{z}}) \equiv  {\tilde{D}}_{\bar{z}}
\phi^{ab}_{T^2} (z, {\bar{z}}) =0
\label{anni45}
\end{eqnarray}
with the following periodicity conditions to be satisfied in going around the
two one-cycles of the torus:
\begin{eqnarray}
\phi^{ab} (z+1,\bar{z}+1)= e^{i\,\chi_1( z,\,\bar{z})}\phi^{ab} (z,\,\bar{z})\qquad
\phi^{ab} (z+U,\bar{z}+\bar{U})= e^{i \,\chi_2 ( z,\,\bar{z})}\phi^{ab} (z,\,\bar{z})
\label{bciwf}
\end{eqnarray}
where
\begin{eqnarray}
\chi_1=\frac{\pi I}{ {\rm Im}U}{\rm Im}(z)~~;~~
\chi_2=\frac{\pi I}{ {\rm Im}U}{\rm Im}(\bar{U}\,z) \,\,.
\label{chiab}
\end{eqnarray}
Remember that we use the following definition of the $\Theta$-function:
\begin{eqnarray}
\Theta\left[\begin{array}{cc} {\alpha  }   \\\beta\end{array}\right]
(  z  |  U ) = \sum_{n} {\rm e}^{2 \pi i \left[ \frac{1}{2}(n + \frac{\alpha}{2})^2 U +
(n + \frac{\alpha}{2}) ( z + \frac{\beta}{2} )   \right]} \,\,.
\label{THETA}
\end{eqnarray}
If $I <0$ the identification of $D_z$ and $D_{\bar{z}}$  with the creation and
annihilation  operators is exchanged; i.e.:
\begin{eqnarray}
  -i{\tilde{D}}_{{{z}}} = \sqrt{\frac{\pi |I|}{U_2}} a~~~~
  -i{\tilde{D}}_{{\bar{z}}} =\sqrt{\frac{\pi |I|}{U_2}} a^\dagger \,\,.
\label{aadaggerI}
\end{eqnarray}
The operator in Eq. (\ref{DD96}) becomes:
\begin{eqnarray}
-{\tilde{D}}_{k}  {\tilde{D}}^{k} =
 \frac{2 \pi |I|}{{\cal{T}}_2}  \left(  2 a^{\dagger} a +1 \right) \equiv
 \frac{2 \pi |I|}{{\cal{T}}_2} \left( 2N +1 \right) \,\,.
\label{ope54}
\end{eqnarray}
The wave functions of the (degenerate) ground state,
are given by:
\begin{eqnarray}
\phi^{ab,n}_{T^2}  =
e^{ \pi i|I| {\bar{z}} \frac{ {\rm Im}{\bar z} }{ {\rm Im} U }}\,
\Theta\left[\begin{array}{cc} \frac{-2n}{I }\\0\end{array}\right]
(I  \bar{z}  | I  \bar{U} )  ~~;~~n  =0\dots |I | -1
\label{wafu6}
\end{eqnarray}
and are determined by requiring them to satisfy the following equation:
\begin{eqnarray}
a \,\, \phi^{ab}_{T^2} (z, {\bar{z}}) \equiv  {\tilde{D}}_{{z}}
\phi^{ab}_{T^2} (z, {\bar{z}}) =0
\label{anni45b}
\end{eqnarray}
and the periodicity conditions in Eqs. (\ref{bciwf}).
In particular, the structure of the phase factor
in Eq. (\ref{wafu6}) is fixed by the Laplace equation (\ref{anni45b}),
while the arguments of the Theta function follow from the boundary conditions
in Eqs. (\ref{bciwf}) where we have used that Eq. (\ref{chiab}) can be
equivalently written as
\begin{eqnarray}
\chi_1=\frac{\pi I}{ {\rm Im}{\bar U}}{\rm Im}(\bar z)~~;~~
\chi_2=\frac{\pi I}{ {\rm Im}{\bar{U}}}{\rm Im}({{U}}\,\bar z) \,\,.
\label{chiabb}
\end{eqnarray}
Eqs. (\ref{DD96}) and (\ref{ope54})  can be immediately generalized
to the torus $T^2 \times T^2
\times T^2$  getting:
\begin{eqnarray}
-  {\tilde{D}}_{k}  {\tilde{D}}^{k} \Longrightarrow   \sum_{r=1}^{3}
 \frac{2 U_{2}^{(r)}}{{\cal{T}}_{2}^{(r)}}
 \left\{  {\tilde{D}}_{{{z}_r}}  ,    {\tilde{D}}_{{\bar{z}}_r } \right\}
=  \sum_{r=1}^{3}
 \frac{2 \pi |I_r | }{{\cal{T}}_{2}^{(r)}} \left( 2N_r +1 \right)
\label{Delta6}
\end{eqnarray}
where the arrow indicates the change from the dimensional variables $x^1 , x^2$
to the variables $z, {\bar{z}}$.
In conclusion, Eq. (\ref{eq88}) (in dimensionless compact coordinates)
can be written as:
\begin{eqnarray}
-  {\tilde{D}}_{k}  {\tilde{D}}^{k} \phi^{ab}_{n } = m_{n}^{2}  \phi^{ab}_{n }
\Longrightarrow
   \sum_{s=1}^{3}
 \frac{2 \pi |I_{s}| }{{\cal{T}}_{2}^{(s)}}
 \left( 2N_s +1 \right)\phi^{ab}_{n}={\hat{m}}^2_n\phi^{ab}_{n}~~;~~m_{n}^{2} =
 \frac{{\hat{m}}^2_{n}}{(2 \pi R)^2}\,\,.
\label{Delta61}
\end{eqnarray}
We go on in considering the eigenvalue equation for the fermions given
in Eq.  (\ref{eq88}).
In particular, we restrict ourselves to the case $T^2 \times T^2 \times T^2$ and decompose the six-dimensional Dirac algebra in the product of three two dimensional representations according to the relation~\footnote{See Ref.~\cite{Pol}.}:
\begin{eqnarray}
\gamma_{(6)}^4=\gamma^1_{(1)}\otimes\sigma^3\otimes \sigma^3~~&;&~~
\gamma_{(6)}^5=\gamma^2_{(1)}\otimes\sigma^3\otimes \sigma^3\nonumber\\
\gamma_{(6)}^6=\mathbb{I}\otimes\gamma^1_{(2)}\otimes \sigma^3~~&;&~~
\gamma_{(6)}^7=\mathbb{I}\otimes\gamma^2_{(2)}\otimes \sigma^3\nonumber\\
\gamma_{(6)}^8=\mathbb{I}\otimes\mathbb{I}\otimes\gamma^1_{(3)}~~&;&~~
\gamma_{(6)}^9=\mathbb{I}\otimes\mathbb{I}\otimes\gamma^2_{(3)}
\label{gam66}
\end{eqnarray}
with:
\begin{eqnarray}
\gamma^2_{(r)}\equiv\sigma^2=\left( \begin{array}{cc}
                          0&-i\\
                          i&0
                        \end{array}\right) ~~:~~  \gamma^1_{(r)}\equiv\sigma^1=\left( \begin{array}{cc}
                          0&1\\
                          1&0
                        \end{array}\right) \,\,.
\label{gam67}
\end{eqnarray}
Correspondingly, the ten dimensional Majorana-Weyl spinors are the
product of a four-dimensional spinor  and
three two-dimensional spinors $\eta_{1}\otimes\eta_{2}\otimes\eta_{3}$. The
ten-dimensional Weyl  condition imposes that these latter
have to be Weyl spinors:
\begin{eqnarray}
i\gamma_{(r)}^1\,\gamma_{(r)}^2\eta_{r}=\pm\eta_{r}\label{bwc} \,\,.
\end{eqnarray}
The Dirac matrices, previously  introduced, satisfy the Clifford algebra with a flat metric.
On the torus $T^2$, in the complex coordinates, the metric is given in the second equation in (\ref{meGG}), and therefore the flat Dirac matrices
has to be multiplied by a suitable vierbein: i.e. $\gamma^i= e^i_{~~I}\gamma^I$
that is given in Eq. (\ref{vierbe2}). From it  we get the Dirac matrices
with a ``curved" index:
\begin{eqnarray}
\gamma^z_{(r)}=e^z_{~I}\gamma^I_{(r)}=\sqrt{\frac{U_2^{(r)}}{{\cal T}_2^{(r)}}}\left(\begin{array}{cc}
                                 0&2\\
                                 0&0
                                 \end{array}\right)~~;~~\gamma^{\bar{z}}_{(r)}=e^{\bar{z}}_{~I}\gamma^I_{(r)}=
                                 \sqrt{\frac{U_2^{(r)}}{{\cal T}_2^{(r)}}}\left(\begin{array}{cc}
                                 0&0\\
                                 2&0
                                 \end{array}\right)
\label{curvega}
\end{eqnarray}
and therefore we can write:
\begin{eqnarray}
\gamma^{z^r}_{(6)}= \mathbb{I}^{\otimes (r-1)}\otimes \gamma^{z}_{(r)}
\otimes (\sigma^3)^{\otimes (3-r)}~~;~~
\gamma^{{\bar{z}}^r}_{(6)}= \mathbb{I}^{\otimes (r-1)}\otimes
\gamma^{{\bar{z}}}_{(r)}\otimes (\sigma^3)^{\otimes (3-r)}
\label{gazr6}
\end{eqnarray}
where $V^{\otimes n}=V\otimes\dots \otimes V$ with $n$ $V$-factors.

Having defined the Dirac matrices,  we go back to the eigenvalue
equation in Eq. (\ref{eq88}) and we square it, getting:
\begin{eqnarray}
\left( - {\tilde{D}}_{i}  {\tilde{D}}^{i} \mathbb{I} - \frac{1}{2} [ \gamma^i , \gamma^j ]
{\tilde{D}}_{i}
{\tilde{D}}_{j} \right) \eta_n = \lambda_{n}^{2} \eta_n \,\,.
\label{ferei}
\end{eqnarray}
For a single torus the second term in the l.h.s. of the previous equation is given by:
\begin{eqnarray}
\frac{1}{2} [ \gamma^i , \gamma^j ]  {\tilde{D}}_{i}  {\tilde{D}}_{j} =
\frac{1}{2(2\pi R)^2} [ \gamma^z , \gamma^{\bar{z}} ]  [{\tilde{D}}_{z},
 {\tilde{D}}_{\bar{z}} ]=
\frac{1}{(2 \pi R)^{2}} \frac{2 \pi I}{T_2} \left( \begin{array}{cc} 1 & 0 \\
                  0 & -1 \end{array} \right)
\label{for52}
\end{eqnarray}
where  Eq. (\ref{commu9I}) has been used.  Eqs. {\ref{for52}  and (\ref{DD96})
 we can put the fermionic eigenvalue equation in the following form:
\begin{eqnarray}
&& \frac{1}{(2\pi R)^{2}} \left[ 2 \pi \sum_{r=1}^{3} (2 N_r +1) \frac{|I_r|}{T_{2}^{(r)}} \mathbb{I} \otimes
 \mathbb{I} \otimes  \mathbb{I}  - \frac{2 \pi I_1}{T_{2}^{(1)}} \sigma_3 \otimes
 \mathbb{I} \otimes  \mathbb{I} - \frac{2 \pi I_2}{T_{2}^{(2)}} \mathbb{I} \otimes
 \sigma_3 \otimes
 \mathbb{I}   \right.  \nonumber \\
&& \left.   -  \frac{2 \pi I_3}{T_{2}^{(3)}} \mathbb{I} \otimes \mathbb{I} \otimes
 \sigma_3    \right] \eta^{1}_{n}  \otimes \eta^{2}_{n}  \otimes \eta^{3}_{n}
 = \lambda_{n}^{2} \eta^{1}_{n}  \otimes \eta^{2}_{n} \otimes \eta^{3}_{n}
\label{eigt4}
\end{eqnarray}
where we have decomposed $\eta_n = \eta^{1}_{n} \otimes \eta^{2}_{n} \otimes
\eta^{3}_{n}$.  This equation shows that, for arbitrary signs of $I_r$ ($r=1,2,3$),
there is always a unique zero mode that is a chiral fermion. In particular, if $I_{1,2,3}$
are all positive, then all three wave functions $\eta^{(1,2,3)}$
will have positive chirality: $\sigma_3 \eta = \eta$. Since the original
ten-dimensional fermion is a Weyl fermion with chirality $\chi_{10}$, the
four-dimensional chirality $\chi_4$ will be equal to
$\chi_4 = \chi_{10} \chi_1 \chi_2 \chi_3 $ where $\chi_r$ ($r=1,2,3$) is the chirality
on the $r$-th torus.

Since the zero mode eigenfunction on $T^2 \times T^2 \times T^2$ is the product
of the zero mode eigenfunctions on each torus $T^2$, we will limit ourselves to
the Dirac equation on the torus $T^2$:
\begin{eqnarray}
\left( \gamma^z_{(r)}\tilde{D}_{z^r}+
\gamma^{\bar{z}}_{(r)}\tilde{D}_{\bar{z}^r}\right)
\eta^{ab}_{r}(z^r,\,\bar{z}^r)=0
\label{Di45}
\end{eqnarray}
where we have omitted the index $0$ to simplify the notation,
and it is satisfied when
\begin{eqnarray}
\left( \gamma^z_{(r)}\tilde{D}_{z^r}+\gamma^{\bar{z}}_{(r)}
\tilde{D}_{\bar{z}^r}\right) \eta^{ab}_{r}(z^z,\,\bar{z}^r)
= 2\sqrt{\frac{U_2}{{\cal T}_2}} \left( \begin{array}{cc}
                                                0&\tilde{D}_{z^r}\\
                                            \tilde{D}_{\bar{z}^r} & 0
                                             \end{array}\right)\left(\begin{array}{c}
                                             \eta^{ab}_{r,+}\\\eta^{ab}_{r,-}
                                             \end{array}\right)=0
\label{D46}
\end{eqnarray}
with  $\tilde{D}$  given in Eq. (\ref{aadagger}).
The Weyl condition written in Eq. (\ref{bwc}) imposes that the spinor has to be of the form:
\begin{eqnarray}
\eta_{r,+}= \left(\begin{array}{c}
\eta^{ab}_{r,+}\\0
\end{array}\right) ~~;~~\eta_{r,-}= \left(\begin{array}{c}
0\\\eta^{ab}_{r,-}
\end{array}\right)\label{internalf}
\end{eqnarray}
with $\eta_{+}$ and $\eta_{-}$ spinors with opposite chirality.
By using again Eq. (\ref{aadagger}), we have that the solution of  Eq. (\ref{D46}) imposes:
\begin{eqnarray}
a_{(r)}\,\eta^{ab}_{r,+}\equiv -i \tilde{D}_{\bar{z}^r}\,\eta^{ab}_{r,+}=0 \qquad I_r> 0
\label{annicrea}
\end{eqnarray}
while for $I_r<0$, as previously discussed, the role of creation and
annihilation operators is exchanged and we have:
\begin{eqnarray}
a_{(r)}\,\eta^{ab}_{r,-}\equiv -i \tilde{D}_{{z}^r}\,\eta^{ab}_{r,-}=0 \qquad I_r< 0 \,\,.
\label{annicrea1}
\end{eqnarray}
The previous equations coincide with those for the bosonic degrees of
freedom (Eqs. \ref{anni45} and \ref{anni45b})
and thus the solutions exactly coincide with the ones in Eq. (\ref{wavefu6})
\begin{eqnarray}
\eta^{ab,n_r}_{r,+}=\phi_{r,+}^{ab,n_r}~~;~~
\eta^{ab,n_r}_{r,-}=\phi_{r,-}^{ab,n_r}
\label{identit}
\end{eqnarray}
with $\eta^{ab}_{r,-} =(\eta^{ba}_{r,+})^\dag$. In particular, if $I>0$ ($I<0$),
then the spinor has positive (negative) chirality because the spinor with the
opposite chirality has a wave-function that diverges for large values of $\mbox{Im} z$.

In the last part of this Appendix we extend  the previous analysis
to the case of the fermions of the ${\cal N}=2$ hypermultiplet. Such fermions
appear  in our model when one of the three tori, for example the third torus,
is not magnetized, i.e. $I_3=0$. In this case, the equations  of motion
(\ref{Di45}) for the lowest massless components of the mode expansion, are
along the first two tori analogously to the ${\cal N}=1$ case,
while on the third torus the covariant derivative becomes the normal
derivative. Also the boundary conditions are unchanged on the first two tori,
while on the third one they become just the periodicity conditions of
the wave-function when translated along the two one-cycles of $T^2$.
These simple considerations allow us to immediately  write down
the compact wave functions of the massless hypermultiplet
fermions. They coincide, along the first two tori, with the ones of
the chiral fermions written for example in Eq. (\ref{internalf}), while
are  constant spinors along the third torus. In particular, the
condition (\ref{bwc}) implies that
the constant spinor has to be a Weyl spinor. Depending on its chirality, we
have two different solutions for the internal
wave-function which have opposite six-dimensional chirality:
\begin{eqnarray}
\eta_\alpha=\eta_{1,\,\pm}^{}\otimes\eta_{2,\,\pm}^{}\otimes
\epsilon_{\alpha} \qquad \alpha=\{ \uparrow,\,\downarrow\} ~~;~~
{\epsilon}_{\uparrow}=\left( \begin{array}{c} 1\\0\end{array}\right)~~{\epsilon}_{\downarrow}=\left( \begin{array}{c} 0\\1\end{array}\right)
\label{iwfd}
\end{eqnarray}
where we have suppressed the  index  labeling the mode expansion and the
upper and lower signs can be independently chosen. The lowest
massless fermionic state is now degenerate and, having both the
ten-dimensional fermion and the internal wave-function  a definite chirality, we have
two four-dimensional fermions with opposite chirality.  as expected.
The full ten-dimensional wave-function can be written as follows:
\begin{eqnarray}
\Psi_\alpha (x,y)=N_{\psi_\alpha}\,\psi_{\alpha} (x^{\mu}) \otimes
\eta_{\beta} (y^i) \qquad \alpha, \beta =\{ \uparrow,\,\downarrow\} \,\,.
\label{wavef}
\end{eqnarray}

\section{Evaluating the Yukawa couplings}
\label{C}

In this Appendix we give some details of the evaluation of the Yukawa couplings
discussed in Sect. \ref{YUKA} for the chiral multiplet. In particular, we will show
how  Eqs. (\ref{YY}), (\ref{YY2}) and (\ref{YUK}) can be obtained
starting from Eq. (\ref{S31}) and considering only the zero modes.
Let us concentrate our attention on the case in which the massless
scalar is along the first torus (namely the condition in Eq. (\ref{sucon4})is satisfied).
Eq. (\ref{S31}) becomes
\begin{eqnarray}
S^{\Phi}_{3} &=&\frac{1}{2g^2}\sqrt{\frac{{\cal T}_2^{(1)}}{2U_2^{(1)}}}
\int d^4 x \sqrt{G_{4}}
\prod_{r=1}^{3} \left[ (2 \pi R)^2 \int d^2 z_r
\sqrt{G^r}      \right]\nonumber\\
&& \times \left[\bar{\psi}^{ca}\,\gamma^5_{(4)}\varphi^{ab}_{{1,-}}\,
\psi^{bc}\otimes (\eta^{ac})^\dag
\gamma^{z^1}_{(6)} \phi^{ab}_{1,+}
\phi^{ab}_{2,{sign I^{ab}_2}}\phi^{ab}_{3,{sign I^{ab}_3}} \eta^{bc}\right.\nonumber\\
&+&\left. \bar{\psi}^{ca} \,\gamma^5_{(4)}
\varphi_{1,+}^{ab}\, \psi^{bc}
\otimes(\eta^{ac})^\dag
\gamma^{{\bar{z}}^1}_{(6)} \phi^{ab}_{1,-}\phi^{ab}_{2,{sign I^{ab}_2}}
\phi^{ab}_{3,{sign I^{ab}_3}}\eta^{bc} \right.\nonumber\\
&-&\bar{\psi}^{ca}\,\gamma^5_{(4)}\varphi^{bc}_{{1,-}}\,
\psi^{ab}\otimes (\eta^{ac})^\dag
\gamma^{z^1}_{(6)} \phi^{bc}_{1,+}
\phi^{bc}_{2,{sign I^{bc}_2}}\phi^{bc}_{3,{sign I^{bc}_3}} \eta^{ab}\nonumber\\
&-&\left. \bar{\psi}^{ca} \,\gamma^5_{(4)}
\varphi_{1,+}^{bc}\, \psi^{ab}
\otimes(\eta^{ac})^\dag
\gamma^{{\bar{z}}^1}_{(6)} \phi^{bc}_{1,-}\phi^{bc}_{2,{sign I^{bc}_2}}
\phi^{bc}_{3,{sign I^{bc}_3}}\eta^{ab} \right]
\label{S1}
\end{eqnarray}
where we have omitted the index $0$ to simplify the notation and inserted  the indexes $a,b,c$ in order to distinguish  the different brane magnetizations.
According to the choice of the Chern classes signs, only one of the four terms in Eq. (\ref{S1})
corresponds to the Yukawa coupling  of the massless  boson with the two fermions.
Thus one has to compute one of the following integrals over the compact space
\begin{eqnarray}
&&\frac{1}{2g^2}\sqrt{\frac{{\cal T}_2^{(1)}}{2U_2^{(1)}}}
\prod_{r=1}^{3} \left[ (2 \pi R)^2 \int d^2 z_r
\sqrt{G^r}      \right]   \nonumber\\
&\times &  \phi^{ab}_{1,+}( \eta^{ac\,\dag}_1
\gamma^{{z}}_{1}\eta_1^{bc})\otimes\phi^{ab}_{2,{sign I^{ab}_2}}
(\eta^{ac\,\dag}_2
\sigma_{3}\eta_2^{bc})\otimes\phi^{ab}_{3,{sign I^{ab}_3}}
(\eta^{ac\,\dag}_3
\sigma_{3}\eta_3^{bc})  \nonumber\\
&=&  \frac{1}{\sqrt{2}g^2}\prod_{r=1}^3
\left[ (2 \pi R)^2 \int d^2 z_r
\sqrt{G^r}      \right]   \nonumber\\
&\times & (\eta^{ca}_{1,-}\, \phi^{ab}_{1,+}\, \eta^{bc}_{1,-})(sign I^{ac}_2\eta^{ca}_{2,\mp}\,\phi^{ab}_{2,{sign I^{ab}_2}}\, \eta^{bc}_{2,\pm})(sign I^{ac}_3\eta^{ca}_{3,\mp}\, \phi^{ab}_{3,{sign I^{ab}_3}}\,\eta^{bc}_{3,\pm})
\label{sS2}
\end{eqnarray}
for the case $I_1^{ab}>0$ and  $I_1^{bc}, I_1^{ca}<0$,
\begin{eqnarray}
&&\frac{1}{2g^2}\sqrt{\frac{{\cal T}_2^{(1)}}{2U_2^{(1)}}}
\prod_{r=1}^{3} \left[ (2 \pi R)^2 \int d^2 z_r
\sqrt{G^r}      \right]  \nonumber\\
&\times &  \phi^{ab}_{1,-}(\eta^{ac\,\dag}_1
\gamma^{\bar{z}}_{1}\eta_1^{bc})\otimes\phi^{ab}_{2,{sign I^{ab}_2}}
(\eta^{ac\,\dag}_2
\sigma_{3}\eta_2^{bc})\otimes\phi^{ab}_{3,{sign I^{ab}_3}}
(\eta^{ac\,\dag}_3
\sigma_{3}\eta_3^{bc})
\nonumber\\
&=&\frac{1}{\sqrt{2}g^2}
\prod_{r=1}^{3} \left[ (2 \pi R)^2 \int d^2 z_r
\sqrt{G^r}      \right]  \nonumber\\
&\times & (\eta^{ca}_{1,+}\, \phi^{ab}_{1,-}\, \eta^{bc}_{1,+})(sign I^{ac}_2\eta^{ca}_{2,\mp}\,\phi^{ab}_{2,{sign I^{ab}_2}}\, \eta^{bc}_{2,\pm})(sign I^{ac}_3\eta^{ca}_{3,\mp}\, \phi^{ab}_{3,{sign I^{ab}_3}}\,\eta^{bc}_{3,\pm})
\label{sS1}
\end{eqnarray}
for the case $I_1^{ab}<0$ and  $I_1^{bc}, I_1^{ca}>0$,
\begin{eqnarray}
&-&\frac{1}{\sqrt{2}g^2}
\prod_{r=1}^{3} \left[ (2 \pi R)^2 \int d^2 z_r
\sqrt{G^r}      \right]  \nonumber\\
&\times &  (\eta^{ca}_{1,-}\, \phi^{bc}_{1,+}\, \eta^{ab}_{1,-})(sign I^{ac}_2
\eta^{ca}_{2,\mp}\,\phi^{bc}_{2,{sign I^{bc}_2}}\,
\eta^{ab}_{2,\pm})(sign I^{ac}_3\eta^{ca}_{3,\mp}\,
\phi^{bc}_{3,{sign I^{bc}_3}}\,\eta^{ab}_{3,\pm})
\label{sS1c}
\end{eqnarray}
for $I_1^{bc}>0$ and  $I_1^{ab}, I_1^{ca}<0$,
and
\begin{eqnarray}
&& -\frac{1}{\sqrt{2}g^2}
\prod_{r=1}^{3} \left[ (2 \pi R)^2 \int d^2 z_r
\sqrt{G^r}      \right]  \nonumber\\
&&
\times (\eta^{ca}_{1,+}\, \phi^{bc}_{1,-}\,
\eta^{ab}_{1,+})(sign I^{ac}_2\eta^{ca}_{2,\mp}\,\phi^{bc}_{2,{sign I^{bc}_2}}\,
\eta^{ab}_{2,\pm})(sign I^{ac}_3\eta^{ca}_{3,\mp}\,
\phi^{bc}_{3,{sign I^{bc}_3}}\,\eta^{ab}_{3,\pm})
\label{sS1b}
\end{eqnarray}
for
$I_1^{bc}<0$ and  $I_1^{ab}, I_1^{ca}>0$.
The cases in which $I^{ca}_1>0$ and  $I_1^{ab}, I_1^{bc}<0$
and the one in which $I^{ca}_1<0$ and  $I_1^{ab}, I_1^{bc}>0$
can be obtained from Eqs. (\ref{sS1c}) and (\ref{sS1b}) respectively,
by changing the indices $ca$ with $bc$.
Notice that, in order to get a non-vanishing expression,
 the two fermionic components
of the internal wave function along the second and the third
torus need to have  opposite chiralites.

With the choice of the sign for the Chern classes along the first torus given in Eq. (\ref{sceltaIa}),
the internal wave function associated with
the bosonic zero mode solution is
the first one  in Eq. (\ref{wavefu6}), and thus only the first term  in Eq. (\ref{S1})
contributes. In this case to determine the  Yukawa coupling one has to evaluate the integral in Eq. (\ref{sS2}).

In order to write a general expression of the Yukawa couplings which holds
for each of the previous choices of the Chern classes signs,
one can introduce a factor
$\sigma=sign (I_1^{bc}I_1^{ca}I^{ac}_2I^{ac}_3)=\pm 1$ which encodes the signs $sign (I^{ac}_2)$ and $sign (I^{ac}_3)$, relative
to the second and the third torus and the overall sign of the coupling which is $+$ in the case of Eqs. (\ref{sS2}-\ref{sS1})
and $-$ in the case of Eqs. (\ref{sS1c}-\ref{sS1b}).

Substituting Eqs. (\ref{wavefu6})) in the
previous expression, we end up with the following product of
overlap integrals of three
$\Theta$-functions
\begin{eqnarray}
Y=\frac{1}{g^2}\sigma\prod_{r=1}^3  \int_{T^2_r}
d^2z_{r} \sqrt{G^r}\phi^{ab,l_r}_{sign I^{ab}_r}\,
\phi^{ca,n_r}_{sign I^{ca}_r} \phi^{bc,m_r}_{sign I^{bc}_r}.\label{yukw} \,\,.
\end{eqnarray}
Let us first restrict ourselves to the case of the first torus $T^2$. In
order to calculate the previous integral one has to use the
addition formula for the $\Theta$-functions\cite{mum}
\begin{eqnarray}
&&\Theta\left[ \begin{array}{c} \frac{2a}{n_1}\\ 0 \end{array} \right]
(z_1|n_1 \Omega) \Theta\left[ \begin{array}{c} \frac{2b}{n_2}\\ 0\end{array} \right]
(z_2|n_2\Omega)= \sum_{ d\in Z_{(n_1+n_2)} }
\Theta\left[ \begin{array}{c} \frac{2(n_1 d + a + b)}{{n}_1+n_2}\\ 0\end{array}\right]
(z_1+z_2|(n_1+n_2)\Omega)
 \nonumber\\
&&\times\Theta\left[ \begin{array}{c} \frac{2(n_1n_2 d +n_2 a -n_1 b)}{n_1n_2(n_1+n_2)}\\ 0\end{array}\right]
(n_2z_1-n_1z_2|n_1n_2(n_1+n_2)\Omega)
 \label{addizione}
 \end{eqnarray}
 where $Z_{(n_1+n_2)}$ indicates the set of the integer numbers modulo $(n_1+n_2)$.
In our example, being $n_1 \equiv I^{ca}, n_2 \equiv I^{bc}<0$ and $I^{ab}>0$
with $I^{bc} + I^{ca} +I^{ab} =0$, we have
\begin{eqnarray}
&&\phi^{ {ca},n}_-(\bar z) \phi^{{bc},m}_-(\bar z) =
e^{i\pi I^{ab}{\bar z} \frac{ {\rm Im}{\bar z}}{ {\rm Im} U} }
\sum_{d\in \mathbb{Z}_{I^{ca}+I^{bc}}}
\Theta \left[\begin{array}{c} \frac{2(dI^{ca}-n-m)}{I^{ba}}\\ 0\end{array}
\right] (I^{ba}{\bar z}|I^{ba}{\bar U})\nonumber\\
&&\times \Theta \left[ \begin{array}{c}
\frac{2(dI^{bc} I^{ca}-nI^{bc}+mI^{ca})}{I^{ca}I^{bc}I^{ba}}\\
0\end{array}\right] (0|I^{bc}I^{ca}I^{ba}{\bar U}) \label{131}.
\end{eqnarray}
Let us focus  on the terms that depend on $z$ and $\bar{z}$
and leave for a moment aside the last term in Eq. (\ref{131}).
These terms in fact contribute to the integral in Eq. (\ref{yukw}) on
the first torus $T^2$. For each value of the index $d$
in the sum  in Eq. ({\ref{131}),  one has to evaluate the following integral:
 \begin{eqnarray}
\int d^2 z e^{i\pi I^{ab}{\bar z} \frac{{\rm Im}{\bar z}}{{\rm
Im} U}}  \Theta \left[\begin{array}{c}
\frac{2(dI^{ac}+m+n)}{I^{ab}}\\0 \end{array}\right] (I^{ba}{\bar
z}|I^{ba}{\bar U}) e^{i\pi I^{ab}{z} \frac{{\rm Im}{z}}{{\rm Im}
U}} \Theta \left[\begin{array}{c} \frac{2l}{I^{ab}}\\
0\end{array}\right] (I^{ab}z|I^{ab}U) \label{eqford}\,\,.
\end{eqnarray}
By defining:
\begin{eqnarray}
z \equiv x+Uy~~~;~~~0 \leq x\leq 1~~;~~~0 \leq y \leq 1~~;~~U\equiv U_1 + i U_2
\label{zxy}
\end{eqnarray}
the previous integral becomes:
\[
T_2\int_{0}^{1} dx \int_{0}^{1} dy e^{i \pi I^{ab}(U-\bar{U}) y^2}  \sum_{q, q'=-\infty}^\infty
e^{i\pi I^{ab}\left[ \left(q' + \frac{l}{I^{ab}} \right)^2 U -
 \left(q+\frac{dI^{ac}+m+n}{I^{ab}} \right)^2{\bar{U}} \right]}\]
\begin{eqnarray}
\times  e^{2 \pi i I^{ab}\left[ \left(q' + \frac{l}{I^{ab}} \right) -   \left(q+\frac{dI^{ac}+m+n}{I^{ab}} \right) \right] x}
\,\, e^{2i \pi I^{ab}\left[ \left(q' +  \frac{l}{I^{ab}}\right)U -  \left(q+ \frac{dI^{ac}+m+n}{I^{ab}} \right)\bar{U}  \right] y} \,\,.
\label{norm78}
\end{eqnarray}
The integral over $x$ can be easily performed and, in order to get a non-zero result,  one has to impose  the following relation:
\begin{eqnarray}
q+ \frac{dI^{ac}+m+n}{I^{ab}} = q' + \frac{l}{I^{ab}}.
\label{rel72}
\end{eqnarray}
The integral over $y$ can be rewritten as follows:
\begin{eqnarray}
\int_{0}^{1} \!\!\!dy e^{- 2 \pi I^{ab} U_2 y^2} \sum_{q'} e^{- 2 \pi I^{ab} \left(q'+ \frac{l}{I^{ab}} \right)^2 U_2}
e^{- 4 \pi I^{ab}  \left( q'+ \frac{l}{I^{ab}} \right)U_2 y }= \!\!\!\int_{0}^1 \! \! \! dy \sum_{q'} e^{-2\pi I^{ab}U_2
\left( y + q' + \frac{l}{I^{ab}} \right)^2} .
\nonumber
\end{eqnarray}
In conclusion, Eq. (\ref{norm78}) is equal to:
\begin{eqnarray}
T_2 \int_{0}^{1}\!\! du\! \! \sum_{q'= - \infty}^{\infty}
e^{-2\pi I^{ab} U_2 \left( u  + q' +  \frac{l}{I^{ab}} \right)^2}=T_2 \int_{- \infty}^{\infty} \! \! du
e^{-2\pi I^{ab} U_2 \left( u  +  \frac{l}{I^{ab}} \right)^2} =\frac{ T_2 }{  (2  I^{ab} U_2)^{1/2} }
\label{eq93}
\end{eqnarray}
where we have used the identity:
\begin{eqnarray}
\int_{0}^{1} du \sum_{n=- \infty}^{\infty} F (n +u) = \int_{-\infty}^{\infty} du F (u)
\label{ide51}
\end{eqnarray}
which trivially follows from:
\begin{eqnarray}
\int_{0}^{1} du \sum_{n=- \infty}^{\infty} F (n +u)=\lim_{A\rightarrow\infty}\sum_{n=-A}^A\int_{n}^{n+1} dx F(x)=\lim_{A\rightarrow\infty}\int_{-A}^Adx F(x).
\end{eqnarray}

Finally, one gets the following result for the integral in Eq. (\ref{eqford}):
\begin{eqnarray}
&&
\int d^2 z e^{i\pi I^{ab}{\bar z} \frac{{\rm Im}{\bar z}}{{\rm Im} U}}
\Theta \left[\begin{array}{c}\frac{2(dI^{ac}+m+n)}{l}\\0 \end{array}\right]
(I^{ba}{\bar z}|I^{ba}{\bar U})
e^{i\pi I^{ab}{z} \frac{{\rm Im}{z}}{{\rm Im} U}}
\Theta \left[\begin{array}{c} \frac{2l}{I^{ab}}\\ 0\end{array}\right]
(I^{ab}\bar{z}|I^{ab}\bar{U})
\nonumber\\
&&=\frac{T_2}{(2I^{ab} U_2)^{1/2}}\delta_{dI^{ac}+m+n;l}\,\,.
\end{eqnarray}
Here the $\delta$-function comes from  the integration
over $x$, that gives a non-vanishing result only if
\begin{eqnarray}
d I^{ac} + m +n - l=k I^{ab}
\end{eqnarray}
which can be equivalently written as
\begin{eqnarray}
d I^{ac} + m +n - l=0
\label{integ}
\end{eqnarray}
using that the integer $l$ is defined modulus $I^{ab}$.
 On the other hand,
as one can see from Eq. (\ref{131}), $d$ is an integer modulus $I^{ba}$.
Therefore the result on the integration on the first torus $T^2$ is non vanishing
only if, given the integers $m,n,l$, a value of $d$ in the interval
$0 \leq d\leq I^{ab} -1$ can be found such that the
quantity in Eq. (\ref{integ}) is an integer.
Then, including the constant term in  $z$ that appears in
the second line of Eq. (\ref{131})
and assuming that there is a value of $d$ such that the quantity
in Eq. (\ref{integ}) is an integer,
one gets the contribution to  the Yukawa coupling coming from
 the first torus $T^2$:
\begin{eqnarray}
Y=\frac{\sigma}{g^2} \frac{T_2}{(2I^{ab} U_2)^{1/2}}
\Theta \left[ \begin{array}{c}
\frac{2m}{I^{ca}I^{bc}}-\frac{2l}{I^{ca}I^{ba}}\\ 0\end{array}\right]
(0|I^{bc}I^{ca}I^{ba}{\bar U})\,\,.
\label{t2pri}
\end{eqnarray}
The charactheristc of the $\Theta$ function can be written in a
more general way, which is valid for each value of the
Chern classess,
as follows \cite{0404229}:
\begin{eqnarray}
&&\frac{2m}{I^{ca}I^{bc}}-\frac{2l}{I^{ca}I^{ba}}=\frac{2}{I^{ca}}\left(
\frac{m}{I^{bc}}+\frac{l}{I^{ab}}
\right)=\frac{2}{I^{ca}}\left(
\frac{m'I^{ab}}{I^{bc}}+\frac{l'I^{bc}}{I^{ab}}
\right)\nonumber\\
&=&
-2\left\{\frac{m'+l'}{I^{ca}}+
\frac{m'}{I^{bc}}+\frac{l'}{I^{ab}}\right\}
=-2\left\{\frac{n'}{I^{ca}}+
\frac{m'}{I^{bc}}+\frac{l'}{I^{ab}}
\right\}
\label{ifin}
\end{eqnarray}
where we have made a ridefinition of the indices $m, l \rightarrow m', l'$ which are still defined modulus $I^{bc}$ and
modulus $I^{ab}$ respectively. Such a redefinition is allowed for $(I^{bc},I^{ab} ,I^{ca})$
relative prime.  Moreover, we have introduced $n'=m'+l'$ which is defined modulus $I^{ca}$.
Then substituting  Eq. (\ref{ifin}) in  Eq (\ref{t2pri}) one gets
\begin{eqnarray}
Y=\frac{\sigma}{g^2} ~ \frac{T_2}{(2I^{ab} U_2)^{1/2}}
\Theta \left[ \begin{array}{c} 2\left(\frac{n'}{I^{ca}}+
\frac{m'}{I^{bc}}+\frac{l'}{I^{ab}}\right)
\\ 0\end{array}\right]
(0|I^{bc}I^{ca}I^{ba}{\bar U})
\label{Yzy}
\end{eqnarray}
where we have omitted the minus sign
in the characteristic of the $\Theta$-function and used the property
$\Theta\left[\begin{array}{c}-a\\0\end{array}\right](0|t)=\Theta\left[\begin{array}{c}a\\0\end{array}\right](0|t)$.

In order to generalize the previous result to the case of the torus $T^2\times T^2\times T^2$ (always performing the choice in Eq. (\ref{sceltaIa})),
we notice  from Eq. (\ref{sS2}) that the following integral has to be computed
both along the second and the third torus:
\begin{eqnarray}
\int_{T^2_r} d^2z_r \sqrt{G^r}
\phi^{ca}_{r,\mp}\, \phi^{ab}_{r,sign I^{ab}} \phi^{bc}_{r,\pm} \,\,.
\label{tori23}
\end{eqnarray}
One has to apply again the addition formula (\ref{addizione})
to $\phi^{ab}_r$ and $\phi^{ca}_r$ if $sign(I^{ab}_rI^{ca}_r)>0$
or to $\phi^{ab}_r$ and $\phi^{bc}_r$ if $sign(I^{ab}_rI^{bc}_r)>0$.
Following the same calculations done in Eqs.
(\ref{131})-(\ref{eq93}) one ends with a normalization factor
\begin{eqnarray}
&&\frac{ T_2^{(r)} }{  (2  |I^{bc}_r| U_2^{(r)})^{1/2} } \rightarrow {\rm for} ~~sign(I^{ab}_rI^{ca}_r)>0\nonumber\\
&&\frac{ T_2^{(r)} }{  (2  |I^{ca}_r| U_2^{(r)})^{1/2} } \rightarrow {\rm for} ~~sign(I^{ab}_rI^{bc}_r)>0 \,\, .
\label{tori23a}
\end{eqnarray}
Notice that with a choice different from the one in (\ref{sceltaIa}), for instance
$I_{1}^{ab}, I_{1}^{ca}<0$ and $I_{1}^{bc}>0$), one should start from  Eq. (\ref{sS1c}) rather than
Eq. (\ref{sS2}) and thus the integral to compute along the second and the
third torus would be
\begin{eqnarray}
\int_{T^2_r} d^2z_r \sqrt{G^r}
\phi^{ca}_{r,\mp}\, \phi^{bc}_{r,sign I^{bc}} \phi^{ab}_{r,\pm}
\label{tori23c}
\end{eqnarray}
and, instead of Eq.(\ref{tori23a}), one would find:
\begin{eqnarray}
&&\frac{ T_2^{(r)} }{  (2  |I^{ab}_r| U_2^{(r)})^{1/2} } \rightarrow {\rm for} ~~sign(I^{bc}_rI^{ca}_r)>0\nonumber\\
&&\frac{ T_2^{(r)} }{  (2  |I^{ca}_r| U_2^{(r)})^{1/2} } \rightarrow {\rm for} ~~sign(I^{ab}_rI^{bc}_r)>0 \,\,.
\label{tori23b}
\end{eqnarray}
Defining
\begin{eqnarray}
\chi^{ab}_r&=&(1+sign (I^{bc}_rI^{ca}_r))/{2}\nonumber\\
\chi^{bc}_r&=&(1+sign (I^{ab}_rI^{ca}_r))/{2}\nonumber\\
\chi^{ca}_r&=&(1+sign (I^{bc}_rI^{ab}_r))/{2}
\label{chidef}
\end{eqnarray}
one can write the two previous results
in a unified way as
\begin{eqnarray}
\frac{T_{2}^{(r)}}{ \left(2 U_{2}^{(r)}|I^{bc}_r|^{\chi^{bc}_r}
 |I^{ab}_r|^{\chi^{ab}_r}|I^{ca}_r|^{\chi^{ca}_r} \right)^{1/2}}
\label{chidef2}
\end{eqnarray}
for all the three tori.

\section{Supersymmetry transformations}
\label{susy}

The action of the ten-dimensional  ${\cal N}=1$ super Yang-Mills, written
in Eq. (\ref{10dimla}),  is invariant under the global supersymmetric
variations~\cite{GSWI}:
\begin{eqnarray}
\delta A_M=\frac{i}{2}\, \bar{\epsilon}\,\Gamma_M \,\lambda~~;~~
\delta\lambda=-\frac{1}{4}\,\Gamma^{MN}\,F_{MN}\,\epsilon
\label{susytra}
\end{eqnarray}
where $\epsilon$ is a ten-dimensional constant spinor.
Starting from the  ten dimensional supersymmetry transformations and performing
the Kaluza-Klein reduction  we can determine the  four-dimensional
supersymmetries  which are preserved in our flux compactification.
This analysis can be carried  in both the twisted and untwisted  sectors.
In the following we restrict our attention only to the twisted sector.
In particular, by implementing in Eq. (\ref{susytra}) the decompositions written in Eq.
(\ref{exp98}), restricting our analysis only to the massless fields
in the bifundamental representation of the gauge group $U(1)_a\times U(1)_b$, using Eq.  (\ref{expa3}) and  the mode expansions given
in Eq. (\ref{exps63}), we can write:
\begin{eqnarray}
N_{\varphi_1}\,\delta[\varphi_{{z}^1}^{ab}(x)\otimes \phi^{ab}_0(y)]=
i\frac{N_\psi}{2} \bar{\epsilon}_4 \gamma_{(4)}^5 \psi^{ab}(x)\otimes G_{z^1\bar{z}^1}\epsilon^\dag_1 \gamma^{\bar{z}^1}\eta_1^{ab}\otimes \epsilon_2^\dag ~\mathbb{I}~ \eta_2\otimes \epsilon_3~ \mathbb{I}~ \eta^{ab}
 \label{susc} \,\,.
\end{eqnarray}
Here, $N_1$ and $N_\psi$ are the  normalization factors that we  have
 introduced in order  to have four dimensional
actions with the correct holomorphic properties, as extensively
discussed in this paper.

Eq. (\ref{susc}) has been obtained by decomposing the
ten-dimensional spinor $\epsilon$ as a product of
a four-dimensional spinor and three two-dimensional ones as follows:
\begin{eqnarray}
\epsilon=\epsilon_{(4)}\otimes\left(\begin{array}{c}
\epsilon_1^+\\\epsilon_1^-\end{array}\right)
\otimes\left(\begin{array}{c}\epsilon_2^+\\\epsilon_2^-\end{array}\right)
\otimes\left(\begin{array}{c}\epsilon_3^+\\\epsilon_3^-\end{array}\right)
\label{epsiab}
\end{eqnarray}
in complete analogy with what we have done for the fermionic field.

The scalar involved in Eq. (\ref{susc}) is massless when the constraint
written in Eq. (\ref{zerom}) is satisfied for $r=1$ and with $I_1^{ab}>0$.
In the following we assume that both these conditions are satisfied.
This means that  the internal total wave-function of the scalar is:
\begin{eqnarray}
\phi^{ab}_0=\phi_{1,\,+}^{ab;\,n^1}\prod_{r=2}^3
\phi_{r,\,sign(I^{ab}_r)}^{ab,\,n^r}~
\label{wavfu}
\end{eqnarray}
where  the wave-functions for each torus  are given in Eq. (\ref{wavefu6}).
 Analogously, the internal two-dimensional spinors  $\eta_r$  (r=1,2,3),
 which come from the decomposition of the six dimensional spinor $\eta$,   have, according
 to the Eq. (\ref{internalf}), positive chirality on the first torus
 and positive or negative chirality on the other two tori, depending on the
sign of $I^{ab}_{r}$ for  $r=2,3$. These considerations allow us to write the susy
transformation, in the following way:
\begin{eqnarray}
N_{\varphi_1} \delta[\varphi_{{z}^1}^{ab}(x)\otimes \phi^{ab}_0(y)]&=&i
\frac{N_\psi}{2}\,\sqrt{\frac{{\cal T}_2^{(1)}}{U_2^{(1)}}} \bar{\epsilon}_4
\gamma_{(4)}^5 \psi^{ab}(x)~(\epsilon_1^-~\eta_{1,\,+}^{ab;\,n^1})~
(\epsilon_2^{sign(I_2^{ab})}\eta_{2,\,sign(I_2^{ab})}^{ab,\,n^2})~\nonumber\\
 &\times&
( \epsilon_3^{sign(I_3^{ab})}\eta_{3,sign(I_3^{ab})}^{ab,\,n^3})
\label{Nde}
 \end{eqnarray}
where  the $\gamma$-matrix written in Eq. (\ref{curvega}) has been used.
This equation shows that, in order to have a non-vanishing  expression,
the constant two-dimensional spinors have to be taken equal to:
$\epsilon_1^-=1$, $\epsilon_{2,\,3}^{sign(I^{ab})}=1$
with all the other components being zero. With this choice
and remembering the relation between the bosonic and fermionic
wave function, written in equation (\ref{identit}),
we have:
\begin{eqnarray}
\phi^{ab}_0=\eta_{1,\,+}^{ab;\,n^1}~\eta_{2,\,sign(I_2^{ab})}^{ab;\,n^2}
~\eta_{3,sign(I_3^{ab})}^{ab;\,n^3}
\label{phi62}
\end{eqnarray}
where now the $\eta$'s are the non-zero  components of the
chiral two-dimensional spinor.
Introducing the scalar field $\varphi_1$ and the relation $N_{\varphi_1}=N_{\psi}/{\sqrt{2\pi}}$, both already defined in Sect. 3,
we can write the four-dimensional supersymmetric variation for the twisted fields as follows:
\begin{eqnarray}
 \delta\varphi^{ab}_1=\frac{i}{2} \bar{\epsilon}_4
 \gamma_{(4)}^5 \psi^{ab}(x)
 \label{sutran}
 \end{eqnarray}
 which explicitly shows that the supersymmetric partner of the fermion
 $\psi$ is the field $\varphi_1$.


\end{document}